\documentclass[apj,tighten,numberedappendix]{emulateapj}
\usepackage{apjfonts}
\defcitealias{fanaroff74}{FR}
\slugcomment{ApJ, in press}  

\newcommand\beq{\begin{equation}}
\newcommand\eeq{\end{equation}}

\shorttitle{Populations of Radio Galaxies}
\shortauthors{Lin et al.}

\newcommand{\allrg}{10,500\ }
\newcommand{\totalrg}{1,040\ }
\begin{document}

\title{On the Populations of Radio Galaxies with Extended Morphology at $\lowercase{z}<0.3$}

\author{
Yen-Ting Lin,\altaffilmark{1,2}
Yue Shen,\altaffilmark{3,2}
Michael A.~Strauss,\altaffilmark{2}
Gordon T.~Richards,\altaffilmark{4}
and
Ragnhild Lunnan\altaffilmark{3,2}
}

\altaffiltext{1}{Institute for the Physics and Mathematics of the Universe, the University of Tokyo, Kashiwa, Chiba, Japan; yen-ting.lin@ipmu.jp}
\altaffiltext{2}{
Princeton University Observatory, 
Princeton, NJ 08544} 
\altaffiltext{3}{
Harvard-Smithsonian Center for Astrophysics, Cambridge, MA 02138}
\altaffiltext{4}{
Department of Physics, Drexel University, Philadelphia, PA 19104}

%%%%%%%%%%%%%%%%%%%%%%%%%%%%%%%%%%%%%%%%%%%%%%
\begin{abstract}

Extended extragalactic radio sources have traditionally been classified into \citeauthor{fanaroff74} (FR) I and II types, based on the ratio $r_s$ of the separation $S$ between the brightest regions on either sides of the host galaxy and the total size $T$ of the radio source ($r_s\equiv S/T$). In this paper we examine the distribution of various physical properties as a function of $r_s$ of \totalrg luminous ($L\gtrsim L_*$) extended radio galaxies (RGs) at $z<0.3$ selected with well-defined criteria from the SDSS, NVSS, and FIRST surveys.
About $2/3$ of the RGs are lobe-dominated (LD), 
%with great majority of fluxes contained in the lobes), 
and $1/3$ have prominent jets. 
If we follow the original definition of the FR types, i.e., a division based solely on $r_s$, FR I and FR II RGs overlap in their host galaxy properties. However, the rare, LD sources with $r_s\gtrsim 0.8$ {\it and} [\ion{O}{3}]$\lambda5007$ line luminosity $>10^6 L_\odot$ are markedly different on average from the rest of the RGs, in the sense that they are hosted in lower mass galaxies, live in relatively sparse environments, and likely have higher accretion rates onto the central supermassive black hole (SMBH). Thus these high emission line luminosity, high $r_s$ LD RGs and the rest of RGs form a well-defined dichotomy.
Motivated by the stark differences in the nuclear emission line properties of the RG subsamples, we suggest that the accretion rate onto the SMBH may play the primary role in creating the different morphologies.
At relatively high accretion rates, the accretion system may produce powerful jets that create the ``classical double'' morphology (roughly corresponding to the LD sources with $r_s\gtrsim 0.8$ and emission lines); at lower accretion rates the jets from a radiatively inefficient accretion flow generate radio lobes without apparent ``hot spots'' at the edge (corresponding to the majority of LD sources). At slightly lower accretion rates {\it and} in galaxies with dense galactic structure, sources with prominent jets result.
It is possible that while the high accretion rate systems could affect sub-Mpc scale environments, the jets from lower accretion rate systems may efficiently suppress activity within the host galaxies.

\end{abstract}


%%%%%%%%%%%%%%%%%%%%%%%%%%%%%%%%%%%%%%%%%%%%%%
\section{Introduction}
\label{sec:intro}

Ever since the seminal work by \citet[][hereafter FR]{fanaroff74}, radio galaxies (RGs) have been broadly categorized into two types according to their large-scale radio morphology (e.g., \citealt{deyoung02}). 
FR proposed a simple binary classification scheme, based on the ratio $r_s$ of the separation between the brightest regions on either sides of the RG and the total size of the radio source. If $r_s>0.5$, the source was considered type II (``edge-brightened''); otherwise the source belonged to type I (``edge-darkened'').
FR found that nearly all 3CR \citep{mackay71} radio sources with power ($P$) 
at 178 MHz greater than $1.3\times 10^{26}$ W/Hz were type II, while nearly all of those weaker than this power were type I. 

%%% 2.5e26*.5^2/.7^2=1.3e26

Subsequent studies have found an overlap in radio power (about two orders of magnitude) for the transition from one type to the other \citep[e.g.,][]{baum89}.
In a series of papers, Owen and collaborators suggested this division between type I and II RGs was a function of the optical luminosity ($L$) of the host galaxies, in the sense that the division was at higher radio power for more luminous galaxies \citep[$P \propto L^{1.8}$;][]{owen89,owen91,owen93,ledlow96}. A corollary is that, at fixed radio power, type I sources are hosted in more optically luminous galaxies than the type IIs. At a given optical luminosity, Owen et al.~suggested that the transition as a function of radio power was quite abrupt.

It was also found that the two types of sources exhibit differences other than the radio power. Type IIs are usually found in less dense environments, and strong emission lines can often be seen in their optical spectra \citep[e.g.,][]{zirbel95,zirbel97,kauffmann08};
type Is, on the other hand, are usually hosted by giant elliptical galaxies, and on average have weaker or no optical nuclear emission lines, which puts them on a different radio power--emission line luminosity correlation than that of FR IIs \citep[][see below]{hine79,zirbel95}.
The cosmological evolution of the two types may also differ significantly (e.g., \citealt{willott01}; but see \citealt{gendre10}).
Collectively, these differences are usually referred to as the FR I/II dichotomy,
which has been the focus of numerous studies \citep[e.g.,][and references therein]{heckman94,baum95, hardcastle07,kauffmann08,baldi10}.

Larger samples and better observations have led to several proposals for refinement/alternatives of radio source classification schemes \citep[e.g.,][]{owen89,leahy93,laing93}.
In particular,
\citet[][hereafter OL89]{owen89} categorized RGs into three types based on their radio morphologies: {\it classical double} (CD) sources are those with ``compact outer hotspots and elongated, diffuse lobes extending from the hotspots back toward the nucleus''; {\it fat double} (FD) objects have ``bright outer rims of radio emission and roundish diffuse radio lobes''; while {\it twin jet} (TJ) RGs ``can be described by symmetric jets originating in the nucleus and extending on both sides'' of the host.
OL89 suggested that the hosts of FD and TJ sources have similar optical properties, and regarded both types, together with the Narrow-Angle Tail (NAT) objects, as FR I (see also \citealt{owen91}). The CD objects were considered to be equivalent to FR II.

In addition to the radio morphologies, RGs have been classified based on their nuclear optical narrow emission line properties (e.g., \citealt{laing94,hardcastle06}).
%Another way to classify RGs is via the nuclear optical narrow emission line properties (e.g., \citealt{laing94,hardcastle06}).
Objects with weak emission lines were generally referred to as low-excitation (LE) RGs, while their counterparts with strong lines were known as high-excitation (HE) RGs.
Such a scheme is believed to better reflect any differences in the central supermassive black hole (SMBH), or/and the physical conditions of the accretion flow onto the central engine \citep[e.g.,][]{kauffmann08}.
On the other hand, classification based on radio morphology is likely more intimately connected to interactions of the jets with the (large-scale) environments \citep[e.g.,][]{deyoung93,kawakatu09}.
%It is important to realize that 
As such,
the correspondence between HE/LE and FR II/I is not perfect; a large fraction of FR IIs have a LE nucleus \citep[e.g.,][]{laing94}, while some FR Is are HE RGs \citep[see][]{heywood07}.
This suggests that {\it a hybrid classification system that incorporates both radio morphology and nuclear emission line activity may perform better in revealing distinct populations of RGs}, which in turn would lead to a fuller understanding of the onset of radio activity, 
%
%This would have important implications for the unification scheme 
as well as the unification
of radio-loud active galactic nuclei \citep[AGNs; e.g.,][]{barthel89,urry95,falcke95, hardcastle06}.

%Numerous studies have discussed the nature and origin of the FR I/II dichotomy \citep[e.g.,][and references therein]{heckman94,baum95,kauffmann08,baldi10}.
%It is possible that differences in the accretion rate onto the central SMBH, in conjunction with the density of the interstellar medium (ISM) or the surrounding environment of the host galaxy, determine the radio morphology \citep[e.g.,][]{baum95,hardcastle07,kawakatu09}, although other mechanisms (e.g., orientation of the jets with respect to the line of sight) probably are at work too.
%A better understanding of the dichotomy 
%
The samples used by many of the previous studies on the FR I/II dichotomy 
were limited by the available data at the time, and thus
often have heterogeneous origins and did not have well-defined selection criteria \citep[e.g.,][]{owen93,zirbel95}.
Furthermore,
it is inevitable that classification of extended radio sources will be subject to the classifiers' experience and preference. Both factors may 
make it difficult to uncover the origin of the dichotomy.

Our primary goal is to understand the origin of the different radio morphologies, which drives us to investigate ways to %``cleanly'' 
separate RGs into distinct populations, by exploring various classification schemes based on radio morphology or/and nuclear activity. 
In addition, we will use $r_s$ as a continuous parametrization of the radio morphology to study the transition from FR I to FR II, and to quantify any bimodality in the physical properties of the host galaxies on the radio power--optical luminosity plane.
We will mainly use our own terminology to refer to the RG populations identified in this paper;
unless specifically noted, we adopt the original definition of FR when we refer to the FR I/II types of extended RGs.

%
%To better investigate the dichotomy, we 
We have constructed a large RG sample with well-defined criteria (see \S\ref{sec:sample}), 
and paid particular attention to the measurements of total size ($T$) of a RG, as well as the separation ($S$) between the highest surface brightness spots (\S\ref{sec:morph}).
From these quantities we define the ratio $r_s$ objectively and reproducibly, and present a classification system that is purely based on radio morphologies (\S\ref{sec:morph}).
%
%
%Our primary goal is to
%study the transition from FR I to FR II, and to quantify any bimodality in the physical properties of the host galaxies, on the radio power--optical luminosity plane (\S\ref{sec:owen}), and as a function of $r_s$ (\S\ref{sec:phys_prop}).
%
In \S\ref{sec:owen} we examine the distribution of RGs on the radio power--optical luminosity plane as a function of $r_s$, while in \S\ref{sec:phys_prop} we study dependences of various physical properties of host galaxies as a function of $r_s$.
Comparisons between radio loud (RL) and radio quiet (RQ)\footnote{We do not use an optical-to-radio luminosity ratio to distinguish between RL and RQ galaxies; by RL we refer to galaxies with 1.4 GHz power $P>10^{23}$ W/Hz, while by RQ we mean galaxies not detected by the NVSS survey (i.e., 20cm flux density $<3$ mJy; see \S\ref{sec:sample}).} galaxies are made in \S\ref{sec:rlvsrq}.
Discussion of the FR dichotomy,
a proposal for an improved classification scheme based on both $r_s$ and nuclear emission line strengths,
comparison of our scheme with those of FR and OL89,
and the possible origin of the morphological differences, 
are presented in \S\ref{sec:disc}.
We summarize our main results in \S\ref{sec:conclusion}.
Possible systematics due to the limitations of the radio data we use, and the uncertainties in the measurement of $r_s$, are discussed in the Appendices.

Most of the radio sources in the local Universe are compact or barely resolved at $\sim 5\arcsec$ resolution (see \S\ref{sec:sample}); this paper concentrates on extended radio sources, and we will study the physical properties of the host galaxies of compact sources in a future publication.

Throughout this paper we adopt a flat $\Lambda$CDM
cosmological model where $\Omega_M=1-\Omega_\Lambda=0.3$
and $H_0=100 h\,{\rm km\,s}^{-1} {\rm Mpc}^{-1}$ with $h=0.70$. As our sources are at $z<0.3$, the small differences between the adopted cosmology and the current best concordance model \citep[e.g.,][]{komatsu10} have a negligible effect on our results.
%

%%%%%%%%%%%%%%%%%%%%%%%%%%%%%%%%%%%%%%%%%%%%%%
\section{The Parent Radio Galaxy Sample}
\label{sec:sample}

We construct the RG sample by cross-matching the Sloan Digital Sky Survey (SDSS; \citealt{york00}) main galaxy spectroscopic sample \citep{strauss02} with the NRAO VLA Sky Survey (NVSS; \citealt{condon98}) and FIRST (Faint Images of the Radio Sky at Twenty-Centimeters; \citealt{becker95}) surveys. 
We start with the DR6 \citep{sdssdr6} version of the NYU Value-Added Galaxy Catalog\footnote{\url{http://sdss.physics.nyu.edu/vagc/}} \citep[VAGC;][]{blanton05}, selecting from the large-scale structure {\tt bbright0} subsample\footnote{This subsample is defined by a constant, extinction-corrected \citet{petrosian76} magnitude limit of $r=17.6$, together with no fiber collision corrections.} 229,379 galaxies with $0.02<z<0.3$ and $M_r^{0.1}\le -21.27$ (i.e., more luminous than the characteristic magnitude $M^*$ in the galaxy luminosity function, \citealt{blanton03b}.  $M_r^{0.1}$ or $r^{0.1}$ denotes the SDSS $r$-band shifted blueward by a factor of 1.1 in wavelength. Similarly, $g^{0.1}$ and $i^{0.1}$ refer to the shifted $g$ and $i$ bands, respectively). 
Selecting RGs with a uniform absolute magnitude limit makes it easy to compute the fraction of RL objects in volume-limited galaxy samples. 
Since the parent VAGC galaxy sample excludes quasars, our sample does not contain radio quasars.
Our survey area covers 6008 deg$^2$.

\begin{figure*}
\epsscale{1.1}
\plottwo{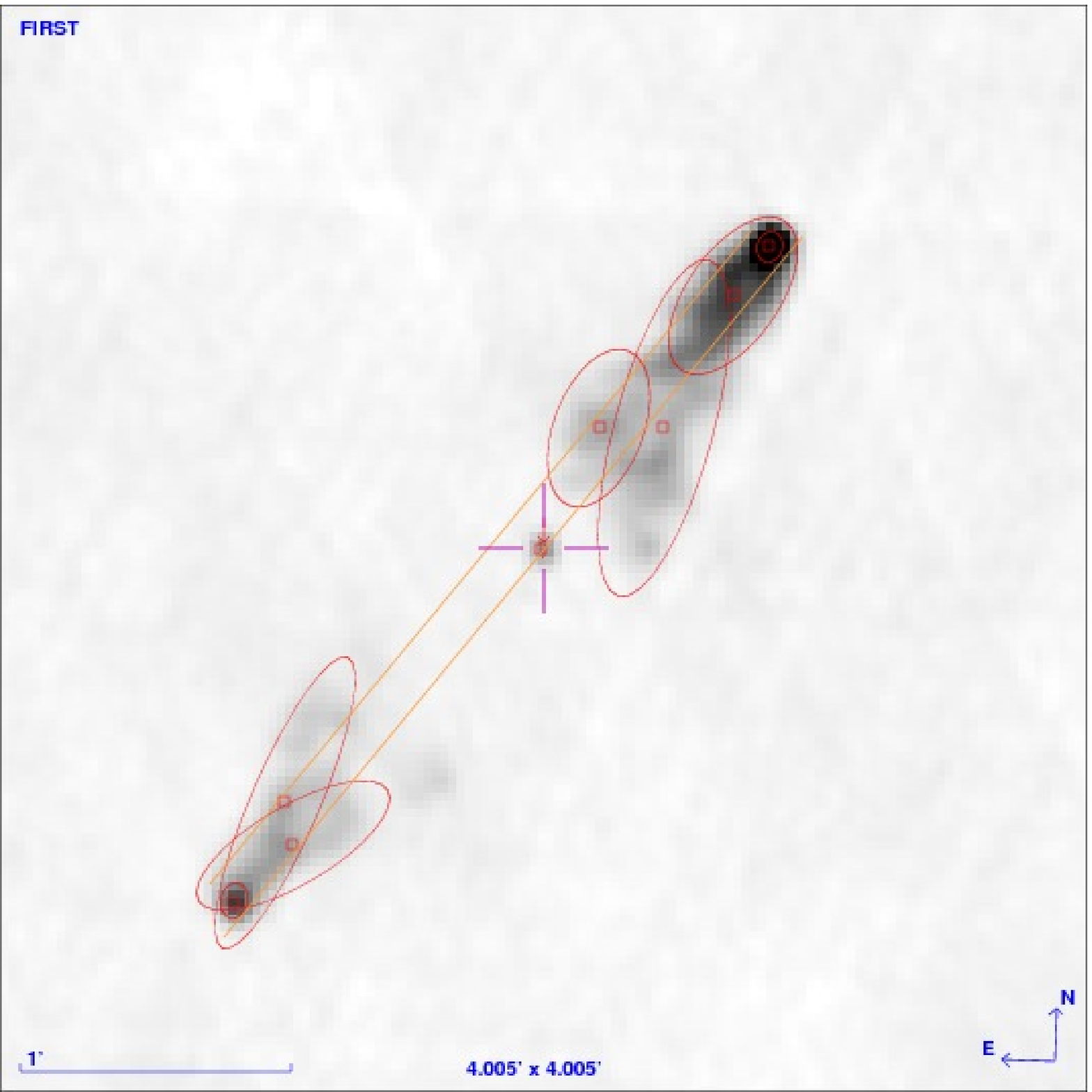}{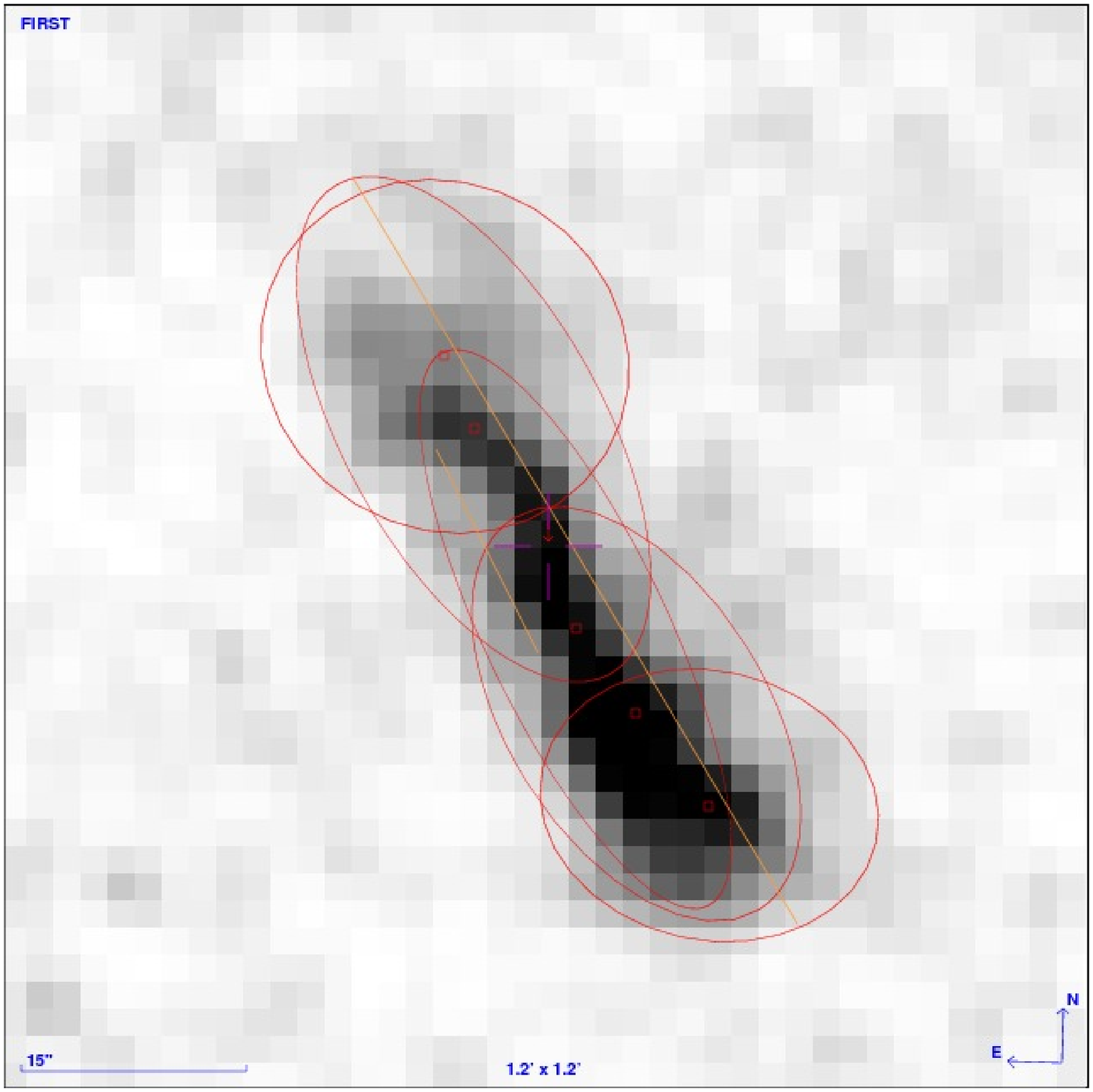}
\vspace{3mm}
\caption{
Illustration of the way we measure the total size $T$ and the separation $S$ between the highest surface brightness (HSB) spots, for class $a$ (i.e., sources with two HSB spots on either side of the galaxy). 
The contrast of these FIRST images are chosen to show various components clearly,
with overlays based on the FIRST source catalog. The small red squares are at the locations of the sources, while the red ellipses represent the best-fit Gaussian models and enclose 95\% of the flux. The cross shows the location of the host galaxy.
In both panels north is up, east is to the left.
{\it Left}: a galaxy at $(\alpha,\delta)=(174.3390,61.3337)$ with $z=0.111$. The image is $4\arcmin \times 4\arcmin$.
A line that connects the two HSB spots (at the edges of the source) and the host galaxy is drawn; $T$ is the separation between the two intersections of the line and the two outermost FIRST ellipses. $S$ is the separation between the HSB spots; in this case, they fall almost along that same line. These two lines are shown in orange in the Figure, offset from each other for clarity. This source has $r_s\equiv S/T=0.93$.
{\it Right}: a galaxy at $(\alpha,\delta)=(161.5181,38.4678)$ with $z=0.127$. The image is $1.2\arcmin \times 1.2\arcmin$.
$T$ is the separation between the intersections of the line that passes most of the centers of the components (i.e., the squares) and the outermost FIRST ellipses (determined in a $\chi^2$-by-eye fashion). $S$ is the separation between the two HSB spots, which are the two components closest to the host galaxy. This source has $r_s=0.26$.
}
\label{fig:example}
\end{figure*}

As the first step, automatic matching to the radio source catalogs is carried out following the prescription of \citet{best05a}. 
In short, if a galaxy has only one NVSS source projected within $3\arcmin$, the pair would be matched depending on their angular separation as well as the properties of the FIRST source(s) (if present) in the vicinity of the galaxy.
If there are at least two NVSS sources within $3\arcmin$ of a galaxy, the matching depends on the spatial distribution and the fluxes of the NVSS and FIRST sources.
We then visually inspect all galaxies with at least one NVSS source within $3\arcmin$ (irrespective of the results of auto-matching), correcting for any mis-match, keeping only RGs with total flux density $f_{1.4}\ge 3$ mJy at 1.4 GHz, and recording their morphological and structural information wherever possible (see \S\ref{sec:morph}). 
The radio flux from the NVSS catalog is in general adopted, 
as fluxes from extended sources may be resolved out by the FIRST survey. 
For complicated sources that are blended/unresolved in the NVSS images, we make use of both NVSS and FIRST data to assign proper fluxes to individual RGs.
More details on the construction of the RG sample will be presented in a future publication where we study the large scale clustering properties of RGs.

Our RG catalog contains about \allrg objects. Of these, \totalrg have extended morphology with roughly aligned lobes/jets; this is the RG sample we study in this paper. The details of the selection and morphology measurement are described in \S\ref{sec:morph}.
Our sample is constructed to be 
complete in radio flux and optical luminosity,
and is not selected against any particular (radio and optical) morphology, which makes it well-suited for investigating the FR I/II dichotomy.
The minimum, mean, and maximum redshift of our sample are 0.027, 0.165, and 0.299, respectively. At $z=0.3$ the resolution of FIRST survey ($5\arcsec$) corresponds to a physical scale of 22 kpc.

Finally, we cross-match all the galaxies in the NYU-VAGC with the DR7 of MPA/JHU-VAGC\footnote{\url{http://www.mpa-garching.mpg.de/SDSS/DR7/}}, which provides continuum-subtracted measurements of emission line strengths \citep{kauffmann03}, star-formation rates \citep{brinchmann04}, and stellar mass \citep{salim07}, among other physical properties.
These auxiliary measurements are used in \S\ref{sec:phys_prop} when we
compare these properties among RGs and between RL and RQ galaxies.

%%%%%%%%%%%%%%%%%%%%%%%%%%%%%%%%%%%%%%%%%%%%%%
\section{Quantification of Radio Source Morphology}
\label{sec:morph}

One of our main goals is to study various properties of RGs and determine if their distribution is bimodal %, as suggested by the FR I/II dichotomy, 
or is continuous. 
Our first task is therefore to define an objective measure (or measures) that allows us to trace the galaxy population smoothly from FR type I-like sources to type II-like ones (as opposed to a sharp and perhaps arbitrary type I vs type II division). With the aid of such a measure, we hope to reduce the subjectiveness inherited in the traditional ways of classification, thus increasing the repeatability of our results by other researchers.

Let us denote the angular separation between the {\it highest} surface brightness (HSB) spots on either sides of the galaxy as $S$, and the total linear size of the radio source as $T$. We follow FR and define $r_s\equiv S/T$ as our primary measure of the morphology of the radio sources. 
We primarily use data from FIRST to measure both $S$ and $T$, except for very large, diffuse sources which become invisible at FIRST resolution.
%We use data from FIRST to measure $S$ (except for very large, diffuse sources which become invisible at FIRST resolution), and use primarily the FIRST image for $T$. 
The reasons for these choices, and the details of the measurements, are described below.

\begin{figure*}
\epsscale{1.1}
\plottwo{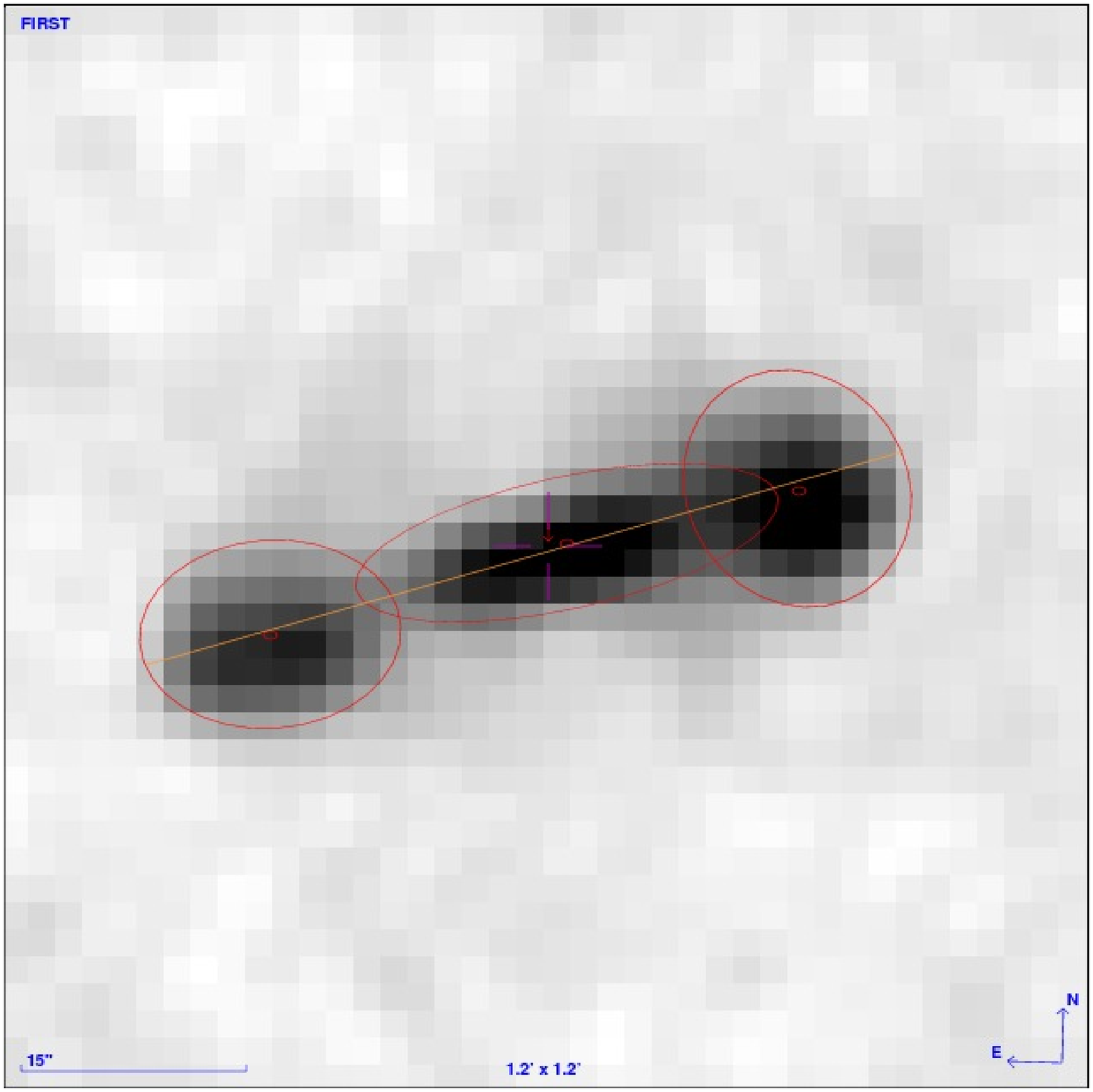}{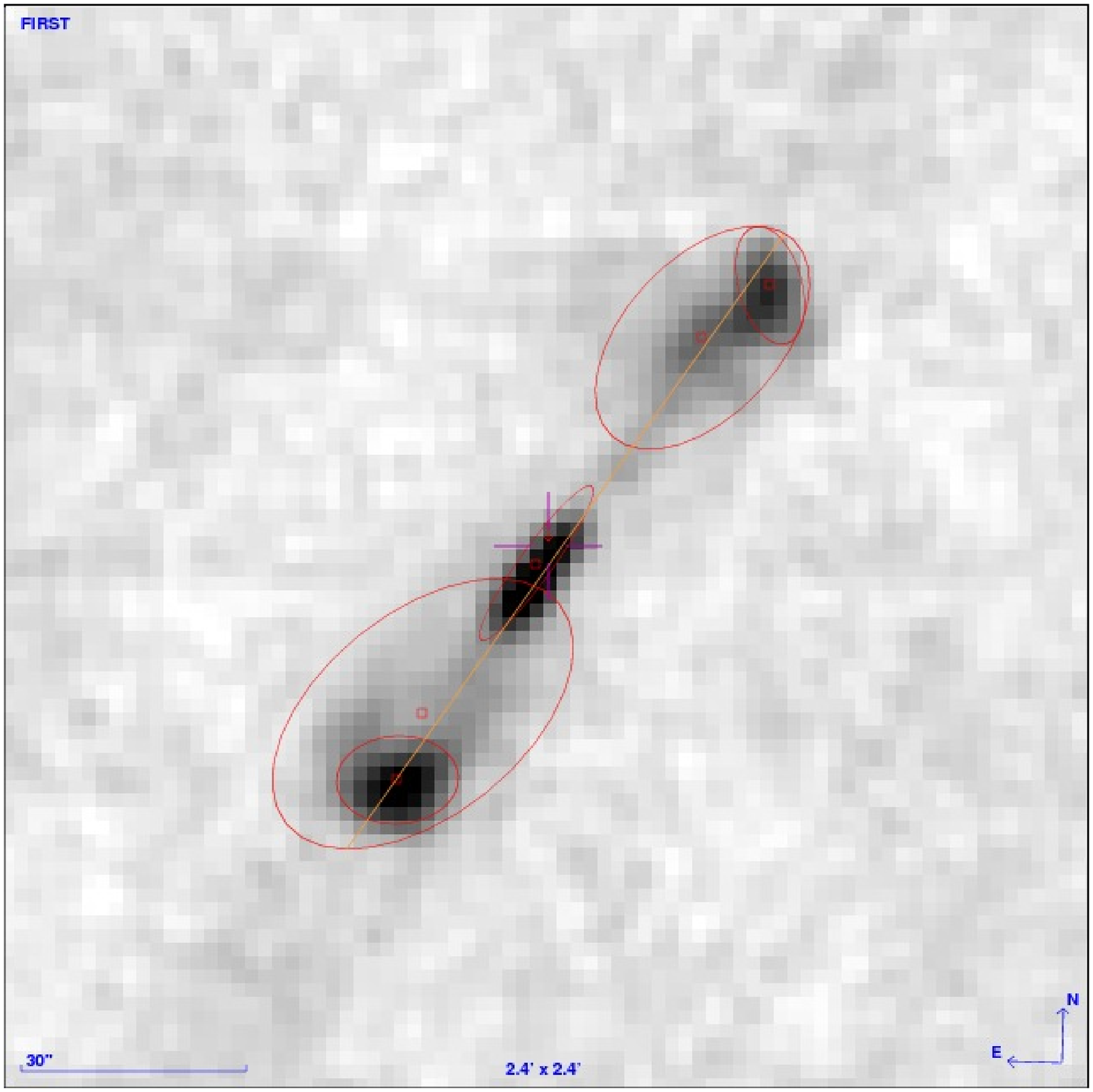}
\vspace{3mm}
\caption{
Similar to Fig.~\ref{fig:example}, but for class $b$, in which the HSB spot is coincident with the galaxy. 
{\it Left}: a galaxy at $(\alpha,\delta)=(112.7111,44.9336)$ with $z=0.072$. The image is $1.2\arcmin \times 1.2\arcmin$.
$T$ is determined in the same way as the examples shown in Fig.~\ref{fig:example}, as indicated by the orange line. In this case we use the major axis of the central FIRST ellipse as $S$. This source has $r_s=0.55$.
{\it Right}: a galaxy at $(\alpha,\delta)=(208.6374,28.2434)$, $z=0.065$, with $r_s=0.25$. The image is $2.4\arcmin \times 2.4\arcmin$.
}
\label{fig:example2}
\end{figure*}

The quantity $S$ depends on the angular resolution of radio maps.
It is possible that a single high surface brightness (SB) spot at low resolution can resolve into several spots at higher resolution, and the identification of the HSB spots can thus change.  
Fifteen of our RGs have been observed by 
a deep 1.4 GHz VLA survey in the SDSS Stripe 82 region
(PI: G.~Richards; see Hodge et al.~2010, in preparation),
at 3 times higher angular resolution (FWHM $\sim 1.5\arcsec$) and twice the sensitivity (rms $\sim 0.07$ mJy/beam) than FIRST (FHWM $5.4\arcsec$, rms $0.15$ mJy/beam). We find that the high SB spots seen in the deeper maps are also identified by FIRST; on average $S$ measured from FIRST and from the deeper survey differ by less than 6\%. We thus conclude that the resolution of FIRST is adequate to identify the highest SB spots for our sample, and our results should be applicable to surveys at arcsec resolution for nearby RGs. More details of this comparison are provided in Appendix \ref{sec:first}.

In addition to the angular resolution, the total size $T$ also depends on the surface brightness limit of the survey.
Although NVSS is shallower than FIRST, it is more sensitive to emission from extended sources with low spatial frequency, which will be absent in the FIRST maps \citep{condon98}.
For most of the sources we measure $T$ using data from FIRST; for 41 very extended, diffuse sources (representing 3\% of all the sources for which we have size measurement), we use NVSS instead.
We discuss in Appendix \ref{sec:totsize} 
the effect of using only NVSS data to measure $T$ on our results.

For the identification of the HSB spots and the measurement of $S$ and $T$, we rely on the elliptical Gaussian fits to the sources provided by both FIRST and NVSS (when needed) surveys, rather than using the atlas images, for these fitted parameters (deconvolved size of major and minor axes, and the position angle) fully account for the synthesized beam properties and are of high fidelity to the actual intensity distribution in the maps \citep{condon98,becker95}. In practice we overlay the FIRST images with ellipses based on the best fit parameters to guide our measurements (see Figures~\ref{fig:example} and \ref{fig:example2} for details). For each source (i.e., individual components of an RG), the ellipse encloses 95\% of the flux, and its semi-major/semi-minor axes are equal to the $2\sigma$ values of the elliptical Gaussian model.

For galaxies associated with only one radio source (at FIRST resolution), or components of a multi-source RG, 
we use the fitted parameters from FIRST to determine whether a source is extended. A point source needs to satisfy the following conditions:
(1) the integrated flux-to-peak flux ratio $f_{\rm int}/f_{\rm peak}<1.2$ and (2) the deconvolved major axis $<5\arcsec$ (e.g., \citealt{becker95,kimball08}).

For every extended RG (containing in most cases at least two FIRST sources\footnote{In this paper we only consider RGs associated with at least two FIRST sources.}) whose lobes/jets are aligned to within $\sim 30^\circ$ (i.e., not strongly bent as in wide/narrow angle tail objects), we measure $T$ and $S$ as follows:
\begin{itemize}
\item $T$: 
{\it (i)} In most of the cases there are two or more FIRST sources associated with an RG; we use the length of the line that passes through the FIRST source locations and intersects with the outermost FIRST source ellipses as $T$ (Fig.~\ref{fig:example}).
{\it (ii)} In the cases where we need to use NVSS for the total size measurement, we use the length of the line that passes through the NVSS/FIRST source locations and intersects with the NVSS source ellipses. If the size of the major axis for a NVSS source in either of the above cases is given as a upper limit, we remove the RG from our sample.
\item $S$: we use the peak flux of the FIRST sources to determine the position of the HSB spots; the separation between such spots on the two sides of the RG is $S$ (e.g., Fig.~\ref{fig:example}). In some cases where one of the lobes is not well detected and modeled by FIRST, we use the FIRST images directly. For about $1/3$ of the sources the HSB spot coincides with the center of the galaxy (Fig.~\ref{fig:example2}), which is likely due to (unresolved) jets; if the spot is extended  and accounts for at least 10\% of the total flux, we use the size of the spot as a measure of $S$. Otherwise we use the separation between the other high SB spots on the two sides of the galaxy as $S$.
\end{itemize}
%

% 1089 = nvss
% 1203 = first
% 1244 = hybrid

From visual inspection, 1,244 RGs appear to be extended (with apparent lobes and/or jets).
We have measured $S$ and $T$ by hand for these objects
using the Aladin sky atlas tool \citep{bonnarel00}.
Based on repeated measurements for about fifty sources, we find our procedure is highly repeatable and the resulting sizes usually agree to within $5\%$, except for very complex sources.
To reduce resolution dependence on our classification scheme, and to remove compact sources (e.g., \citealt{odea98}), we further select a subset of \totalrg sources that have angular size $T>30\arcsec$ and physical size $T_p>40$ kpc for the analysis presented below. Again, our RGs also satisfy $M_r^{0.1}\le -21.27$, $0.02<z<0.3$, and $f_{1.4}\ge 3$ mJy.

\begin{deluxetable*}{cccccc}

\tablecaption{Our Classification Scheme}
%\tablewidth{0pt}

\tablehead{
\colhead{class} & \colhead{HSB\tablenotemark{$\dagger$} spot(s)} & \colhead{central source} & \colhead{$S$} & \colhead{number} & \colhead{fraction (\%)}
}

\startdata
$a$ & opposite sides of RG & not HSB & distance between two HSB spots & 667 & 64.1\\
$b$ & center & extended, $\ge 0.1 f_{\rm tot}$ & size of the HSB spot & 289 & 27.8\\
$c$ & center & extended, $<0.1 f_{\rm tot}$ & distance between secondary HSB spots & \phn38 & \phn3.6\\
$d$ & center & compact, $\ge 0.3 f_{\rm tot}$ & distance between secondary HSB spots & \phn\phn9 & \phn0.9\\
$e$ & center & compact, $< 0.3 f_{\rm tot}$ & distance between secondary HSB spots & \phn37 & \phn3.6
\enddata

\tablenotetext{$\dagger$}{``highest surface brightness''}

\label{tab:classes}

\end{deluxetable*}

Motivated by the above considerations, and
based on the ratio $r_s$, flux, and location of the HSB spot(s) (extended or point-like; coincident with the host galaxy or not), we group the RGs into five classes: 
{\it (a)} There are two HSB spots on opposite sides of the RG;  
{\it (b)} There is a single extended HSB spot coincident with the galaxy, with flux $\ge 0.1 f_{\rm tot}$, where $f_{\rm tot}$ is the total flux from all the components of the RG; 
{\it (c)} The HSB spot coincides with the galaxy, is extended, and its flux is $< 0.1 f_{\rm tot}$; 
{\it (d)} The HSB spot coincides with the galaxy, is unresolved, and its flux is $\ge 0.3 f_{\rm tot}$; 
{\it (e)} The HSB spot coincides with the galaxy, is unresolved, and its flux is $< 0.3 f_{\rm tot}$.
For class {\it b}, $S$ is the size of the HSB spot; for all other classes,
$S$ is the separation between the other high SB spots on both sides of the galaxy.

This classification scheme is devised primarily for the ease of measuring $r_s$. We do not mean to suggest there are five distinct types of RGs.
In Table~\ref{tab:classes} we summarize this scheme, and list the number of sources in each of the classes. The first two classes ($a$ and $b$) account for the majority ($>90\%$) of the extended RGs. We will focus exclusively on these two classes in what follows,
and will examine if such an artificial classification corresponds to any physically different RG types, that is, if there is any fundamental difference between classes $a$ and $b$ (\S\S\ref{sec:phys_prop}, \ref{sec:disc}).

Does $r_s$ have any physical meaning? 
A HSB spot represents a region where the jet is highly dissipative, and thus $r_s$ can be thought of as an indicator of the degree of interaction between the jet and the environment, which depends on the strength and nature of the jet as well as the density of the surrounding medium, among other factors.
For class $a$ sources with $r_s\sim 1$ (e.g., Fig.~\ref{fig:example}, left), the jets can reach the edge of the lobes without much impediment, suggesting either a strong jet or tenuous ambient density, or a combination of both (e.g., \citealt{kawakatu09}). The opposite situation would cause a low $r_s$ (Fig.~\ref{fig:example}, right), or a class $b$ object (Fig.~\ref{fig:example2}).

\begin{figure}
\epsscale{1}
\plotone{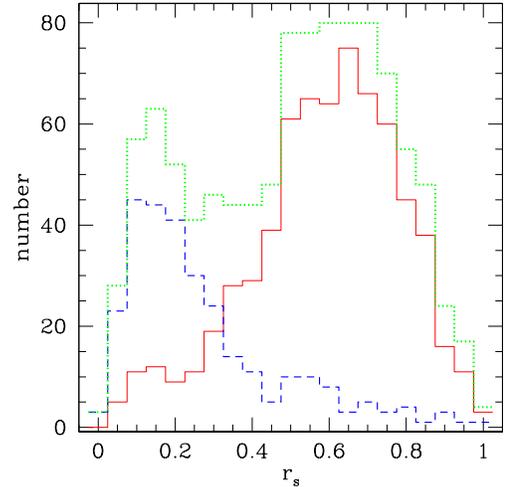}
\vspace{-3mm}
\caption{
Distribution of $r_s$ for all the RGs (green/dotted), and for classes $a$ (red/solid) and $b$ (blue/dashed) separately. 
Classes $c$--$e$ account for about 8\% of the RGs in our sample (see Table~\ref{tab:classes}). The two peaks do {\it not} correspond to the two FR types (e.g., \S\ref{sec:threetypes}).
}
\label{fig:sthist}
\end{figure}

In Fig.~\ref{fig:sthist} we show the distribution of $r_s$ for all \totalrg RGs (green/dotted histogram), and for the classes {\it a} (red/solid histogram) and {\it b} (blue/dashed histogram) separately. 
Considered together (green/dotted histogram), the $r_s$ distribution of RGs seems to be bimodal, with peaks at 0.15 and 0.60, and the division roughly at $r_s\sim 0.3$. 

\begin{figure*}
%\begin{figure}
\epsscale{0.9}
%\epsscale{1.37}
\plotone{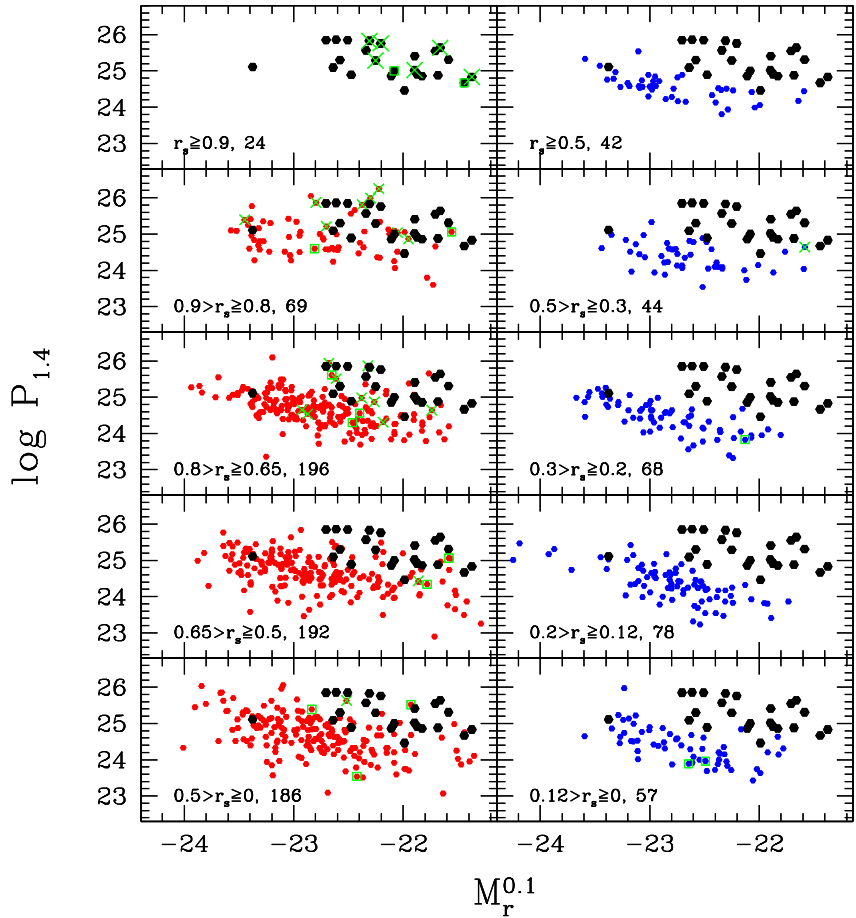}
%\hspace{-5mm}
\vspace{-9mm}
\caption{
Distribution in the radio power--optical magnitude plane for objects in the class $a$ (left panels) and class $b$ (right panels). From top to bottom, the panels show the distribution in bins of decreasing $r_s$. At the lower left corner of each panel we show the $r_s$ range, followed by the number of sources in that range.
The black points (upper left panel) are class $a$ objects with $r_s\ge 0.9$, and are repeated in all other panels for comparison.
There is a systematic shift of the locus of the RGs towards lower radio power, as $r_s$ decreases, for class $a$. 
However, there are class $a$ objects with lower value of $r_s$ in the region occupied by the highest $r_s$ RGs.
This trend is weaker for class $b$.
In each panel, (green) crosses denote sources with Seyfert-like emission line ratios, while squares denote those with LINER-like line ratios (see \S\ref{sec:rlvsrq} and Table~\ref{tab:agn}). 
}
\label{fig:manymp}
\end{figure*}
%\end{figure}

In the original FR scheme, the division was set to $r_s=0.5$, although it is not clear if FR's data showed any bimodality in the distribution of $r_s$.
FR used 1.4 GHz maps with beam size of $23\arcsec \times 23\arcsec {\rm cosec}({\rm dec})$ for their morphology classification. The difference in the division may be due to the resolution of the radio images (see Appendix \ref{sec:totsize}), and/or to the sample used (the 3CR catalog that they used was limited to very high radio flux threshold and consisted of sources at much higher redshifts than our RGs).
Nevertheless, one might be tempted to make a correspondence between the two distributions in Fig.~\ref{fig:sthist} with the two FR types.

Others have built upon the FR scheme and used
more sophisticated criteria to classify the RGs, such as the presence of jets or ``hot spots'' toward the edge of the lobes (e.g., OL89; \citealt{leahy93,gendre08}).
In the analyses presented below, we will seek the best way to distinguish various populations of RGs; in doing so we will find ourselves defining several subsets of class $a$ ($a_{0.9}$, $a_{<0.8}$, $a_{\rm maj}$, and $a_{\rm 0.9,em}$; see Table~\ref{tab:subclasses}).
We will show that, if one sticks with a simple FR-like classification scheme, a division at $r_s\approx 0.8$ (between what we call the $a_{<0.8}$ and $a_{0.9}$ subsets) best separates objects of different physical properties
(for radio data at FIRST-resolution; \S\ref{sec:phys_prop}).
A scheme that works even better is based both on features of the radio morphology ($r_s$) and the nuclear optical emission properties;
we will refer to these as 
the $a_{\rm 0.9,em}$ ($r_s>0.8$ and $L_{\rm OIII}>10^6 L_\odot$) and $a_{\rm maj}$ (the rest of class $a$)  subsets. 
We will argue in \S\ref{sec:threetypes} that the $a_{\rm 0.9,em}$, $a_{\rm maj}$, and $b$ classes represent three populations of extended RGs, which roughly correspond to the CD, FD, and TJ types of OL89, respectively. The correspondence between these and the FR types will be discussed in \S\ref{sec:threetypes}.

\begin{deluxetable}{llrr}

\tablecaption{Subsets of Class $a$}
%\tablewidth{0pt}

\tablehead{
\colhead{subset} & \colhead{definition} & \colhead{number} & \colhead{fraction (\%)}
}

\startdata
$a_{0.9}$ & $r_s>0.8$ (\S\ref{sec:classa}) & 86 & 13\\
$a_{<0.8}$ & $r_s\le0.8$ (\S\ref{sec:classa}) & 581 & 87 \\
$a_{\rm 0.9,em}$ & $r_s>0.8$ and $L_{\rm OIII}>10^6 L_\odot$ (\S\ref{sec:threetypes}) & 40 & 6\\
$a_{\rm maj}$ & non-$a_{\rm 0.9,em}$ class $a$ objects (\S\ref{sec:threetypes}) & 627 & 94
\enddata

%\tablenotetext{$\dagger$}{``highest surface brightness''}

\label{tab:subclasses}

\end{deluxetable}

%86 586 40 632

%%%%%%%%%%%%%%%%%%%%%%%%%%%%%%%%%%%%%%%%%%%%%%
\section{Bimodality in the Radio--Optical Luminosity Plane?}
\label{sec:owen}

We start by analyzing the distribution of RGs in the radio power--optical magnitude plane ($P$--$M$ plane for short). Owen and collaborators suggested that the transition from one FR type to the other is quite abrupt on this plane (\S\ref{sec:intro}; OL89; \citealt{owen91,owen93,ledlow96}).
If this were true, {\it and} if their FR classification was solely based on morphological measures such as $r_s$, we would expect RGs of different values of $r_s$ to occupy distinct regions in the plane.

In Fig.~\ref{fig:manymp} we show the distribution of objects in classes $a$ (left panels) and $b$ (right panels). 
We have assumed a typical radio spectral index of $\alpha = -0.8$ (with the convention that $f\propto \nu^\alpha$) to convert fluxes to power.
From top to bottom, we show the distribution for RGs in bins of decreasing $r_s$. 
Let us first focus on the class $a$. 
The black points (upper left panel) are RGs with $r_s\ge 0.9$, and are shown in all other panels for comparison.
As we will see below, many of these high $r_s$ objects are distinct from the rest of the population in their host properties and environment.
There is a gradual shift of the locus of the RGs towards lower radio power, as $r_s$ decreases.
Even so, the region occupied by these high-$r_s$ RGs is still populated by some RGs with (much) lower value of $r_s$ (e.g., $r_s\sim 0.5$).
Thus no particular region on this plane is inhabited solely by a type of object defined by some specific range of $r_s$.

For class $b$ (right panels), the trend of decreasing mean radio power with decreasing $r_s$ is very weak, but they tend to be lower luminosity than the $r_s\ge 0.9$ class $a$ objects.

\begin{figure}
\epsscale{1}
%\begin{figure*}
\plotone{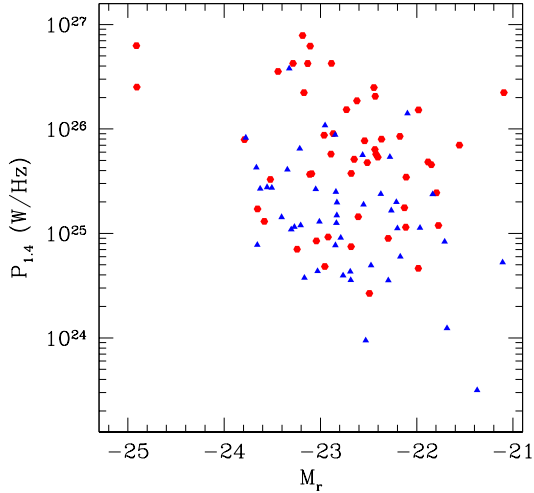}
\vspace{-5mm}
\caption{
Distribution in the radio power--optical magnitude plane of $z<0.3$ FR I (blue/triangle) and FR II (red/circle) RGs from the flux-limited CONFIG sample \citep{gendre10}, matched with the DR7 NYU-VAGC. The morphological classification was done independently by these authors. 
There are 48 (51) type I (II) sources. The upper part (e.g., $P>10^{26}$ W/Hz) of the plane is dominated by type IIs, but there is a non-negligible fraction of type IIs at lower radio power. $M_r$ is the ($k$-corrected) SDSS $r$-band absolute Petrosian magnitude.
}
\label{fig:config}
%\end{figure*}
\end{figure}

These results appear to be in contradiction with those
obtained by Owen and co-workers. 
If we were to assign an $r_s$ value that serves as the FR I/II divide (say $r_s=0.9$), we would not obtain any clean-cut separation for the RGs.
Additional classification criteria, such as special features in the radio source morphology (e.g., hot spots at the edge of the lobes), or/and optical properties of the nuclei, may be required to define two (or maybe more) distinct populations in the $P$--$M$ plane (see \S\ref{sec:mpagain}).

As an independent check, we show in Fig.~\ref{fig:config} the distribution of FR I (blue/triangle) and II (red/circle) RGs in the $P$--$M$ plane, using about 100 RGs at $z<0.3$ from the CONFIG sample \citep{gendre10}.
This is a radio flux-limited sample that was constructed by combining data from both NVSS and FIRST. The morphological classification was carried out by these authors and was based on FIRST and deeper VLA A-array observations.
We cross-match the $z<0.3$ RGs from this sample with the NYU-VAGC DR7 sample to obtain the absolute $r$-band magnitude.
Although FR IIs are on average more powerful than FR Is,
substantial overlap between the two types is readily seen.
This overlap still persists using a volume-limited subsample that consists of 34 FR Is and 28 FR IIs (selected with $z<0.16$ and $M_r\le -21.77$).
In addition, \citet{best09} and \citet{wing10} also noted a substantial overlap between the two FR types in the $P$--$M$ plane.
These results support our notion that in a flux-limited sample, there is not a sharp division among RGs in the $P$--$M$ plane when the classification is made solely based on $r_s$.
Possible causes of the discrepancy between our results and those of Owen et al.~are discussed in \S\ref{sec:mpagain}.

%%%%%%%%%%%%%%%%%%%%%%%%%%%%%%%%%%%%%%%%%%%%%%
\section{Host Galaxy Properties of Extended Radio Sources}
\label{sec:phys_prop}

We now examine the distribution of various physical properties of the galaxies as a function of $r_s$ for classes $a$ and $b$. These include the global properties of the host galaxy, 
derived quantities of the stellar populations, 
emission line properties, 
and properties related to the radio source and environments.

In this section we focus on the observational results, and leave the interpretation to \S\ref{sec:disc}. We discuss the properties of classes $a$ (\S\ref{sec:classa}) and $b$ (\S\ref{sec:classb}) separately. We investigate the environments of the RGs in \S\ref{sec:environment}. A comparison with RQ galaxies whose global properties are matched to the RGs is made in \S\ref{sec:rlvsrq}.

%%%%%%%%%%%%%%%%%%%%%%%
\subsection{Class $a$}
\label{sec:classa}

Throughout this paper we are concerned mainly with the mean behavior of samples, and thus in this section we will examine the medians of various quantities, and their trend with $r_s$.

\begin{figure*}
\epsscale{0.75}
\plotone{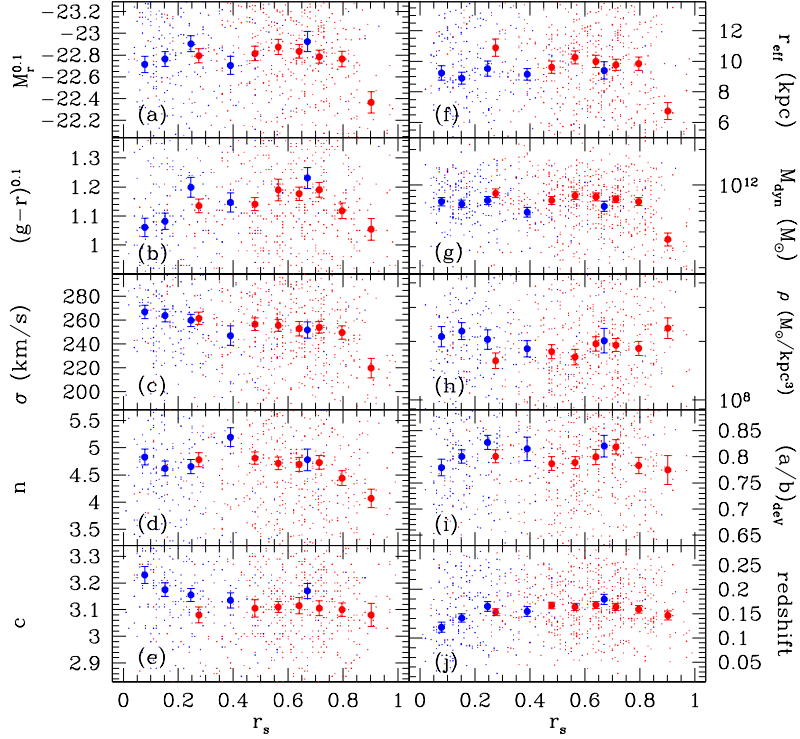}
\vspace{-10mm}
\caption{
Median values of several physical properties of the host galaxies as a function of $r_s$, for classes $a$ (red) and $b$ (blue). 
The errorbars show the uncertainty of the median values. 
Small points are individual RGs. 
Panels (a) to (j) show the absolute $r^{0.1}$-band magnitude, rest frame $(g-r)^{0.1}$ color, velocity dispersion, Sersic index, concentration, effective radius, dynamical mass, mass density within $r_{\rm eff}$, the axis ratio from the de Vaucouleurs profile, and redshift, respectively.
Trends or changes with $r_s$ are in most cases gradual.
}
\label{fig:gal}
\end{figure*}

In Fig.~\ref{fig:gal} we show the median value and 
its uncertainty\footnote{The uncertainty of the median is taken as $\sigma\sqrt{\pi/2N}$, where $\sigma$ is the standard deviation, and $N$ is the number of RGs (e.g., \citealt{lupton93}).}
for some global properties of the host galaxies as a function of $r_s$, for classes $a$ (red) and $b$ (blue). The small data points represent individual RGs, also color coded according to their classes.
The left panels, from top to bottom, show the absolute $r^{0.1}$-band magnitude $M_r^{0.1}$, rest frame color $(g-r)^{0.1}=M_g^{0.1}-M_r^{0.1}$, stellar velocity dispersion $\sigma$, Sersic index $n$, and concentration $c$. 
The panels on the right show the effective radius $r_{\rm eff}$, dynamical mass $M_{\rm dyn}$,
total mass density $\rho$,
axis ratio $b/a$ for the de Vaucouleurs profile fit to the SB profile, and redshift.
Here $c\equiv r_{90}/r_{50}$ is the ratio of the radii that enclose 90\% and 50\% of the Petrosian fluxes, 
$M_{\rm dyn} \equiv 5 r_{\rm eff} \sigma^2/G$ is a crude estimate of the total mass, and
$\rho \equiv M_{\rm dyn}/(4\pi r_{\rm eff}^3/3)$ represents the (central) mass density.
We follow \citet{graham05} to estimate $r_{\rm eff}$ from $n$, $r_{50}$, and $r_{90}$.
The $r$-band measurements are used for all these photometric quantities. 
Using the SMBH $M_{\rm BH}$--$\sigma$ relation (e.g., from \citealt{tremaine02}), we can estimate the SMBH mass $M_{\rm BH}$; for our sample the range of $M_{\rm BH}$ is $2\times 10^7 - 1.3\times 10^9 M_\odot$, with a median of $2.5\times 10^8 M_\odot$.

We have also examined another measure of the galactic structure, the stellar mass surface density, $\mu \equiv M_{\rm star}/\pi r_{\rm eff}^2$, where $M_{\rm star}$ is the stellar mass (described below), and found that the trend with $r_s$ is similar to that of $\rho$.

We focus on class $a$ in this section.
The color, Sersic index, and concentration, and the SB profile (not shown) of the RGs are consistent with those of radio-quiet early type galaxies.
The numbers of RGs in the seven $r_s$ bins are (from lowest to highest): 118, 106, 102, 105, 101, 90, and 49. 
The median redshift is almost constant for different $r_s$ bins (panel j).
Trends or changes with $r_s$ are in most cases gradual and mild, with the exception of the highest-$r_s$ bin.
Although the distributions of the properties are quite broad (c.f.~the small points in Fig.~\ref{fig:gal}), one can identify two populations, roughly separated at $r_s\approx 0.8$: sources with higher $r_s$ (hereafter the $a_{0.9}$ subset) are less luminous and massive (and thus likely harboring smaller SMBH), are smaller, and have slightly higher density. For the rest (hereafter the $a_{<0.8}$ subset), 
trends with $r_s$ are either weak (e.g., $r_{\rm eff}$, $n$, $\sigma$) or absent.
However, it is important to bear in mind that {\it there is significant overlap in physical properties of RGs grouped by the $r_s$ value}. Some of the $a_{0.9}$ objects are as massive ($\sigma$, $M_{\rm dyn}$) and large ($r_{\rm eff}$) as the $a_{<0.8}$ ones (see \S\ref{sec:threetypes}), and, conversely, some of the latter population share the characteristic properties of the former.

\begin{deluxetable*}{rrrrrrrr}

\tablecaption{Fraction of AGN-like Spectrum}
%\tablewidth{0pt}

\tablehead{
\multicolumn{4}{c}{class $a$} & \multicolumn{4}{c}{class $b$}\\
\cline{1-4} \cline{5-8}
\colhead{$r_s$ range} & \colhead{N\tablenotemark{$\ddagger$}} & \colhead{Seyfert\tablenotemark{$\dagger$}} & \colhead{LINER\tablenotemark{$\dagger$}} & \colhead{$r_s$ range} & \colhead{N\tablenotemark{$\ddagger$}} & \colhead{Seyfert\tablenotemark{$\dagger$}} & \colhead{LINER\tablenotemark{$\dagger$}}
}
\startdata

$0.85-1.00$ & 46 & 0.196\phn(0.015) & 0.043\phn(0.049) & $0.50-1.00$ & 37 & 0\phn(0.005) & 0\phn(0.012) \\
$0.75-0.85$ & 85 & 0.082\phn(0.004) & 0.024\phn(0.021) & $0.30-0.50$ & 41 & 0.024\phn(0.005) & 0\phn(0.021)\\
$0.68-0.75$ & 91 & 0.033\phn(0.002) & 0.022\phn(0.028) & $0.20-0.30$ & 58 & 0\phn(0.002) & 0.017\phn(0.046)\\
$0.61-0.68$ & 95 & 0.053\phn(0.008) & 0.011\phn(0.019) & $0.12-0.20$ & 67 & 0\phn(0.003) & 0\phn(0.030)\\
$0.53-0.61$ & 88 & 0\phn(0.001) & 0.011\phn(0.017) & $0.00-0.12$ & 53 & 0\phn(0.008) & 0.008\phn(0.040)\\
$0.42-0.53$ & 93 & 0\phn(0.003) & 0.022\phn(0.019) & \nodata & \nodata & \nodata & \nodata\\
$0.00-0.42$ & 94 & 0.011\phn(0.004) & 0.021\phn(0.030) & \nodata & \nodata & \nodata & \nodata

\enddata

\tablenotetext{$\ddagger$}{Number of all RGs with good velocity dispersion and spectral line measurements.}

\tablenotetext{$\dagger$}{The numbers in parentheses are for radio-quiet (RQ) galaxies whose properties are matched to the radio-loud (RL) galaxies in the $r_s$ bin. See \S\ref{sec:rlvsrq} for more details in the matching between RL and RQ galaxies.}

\label{tab:agn}

\end{deluxetable*}

\begin{figure*}
\epsscale{0.75}
\plotone{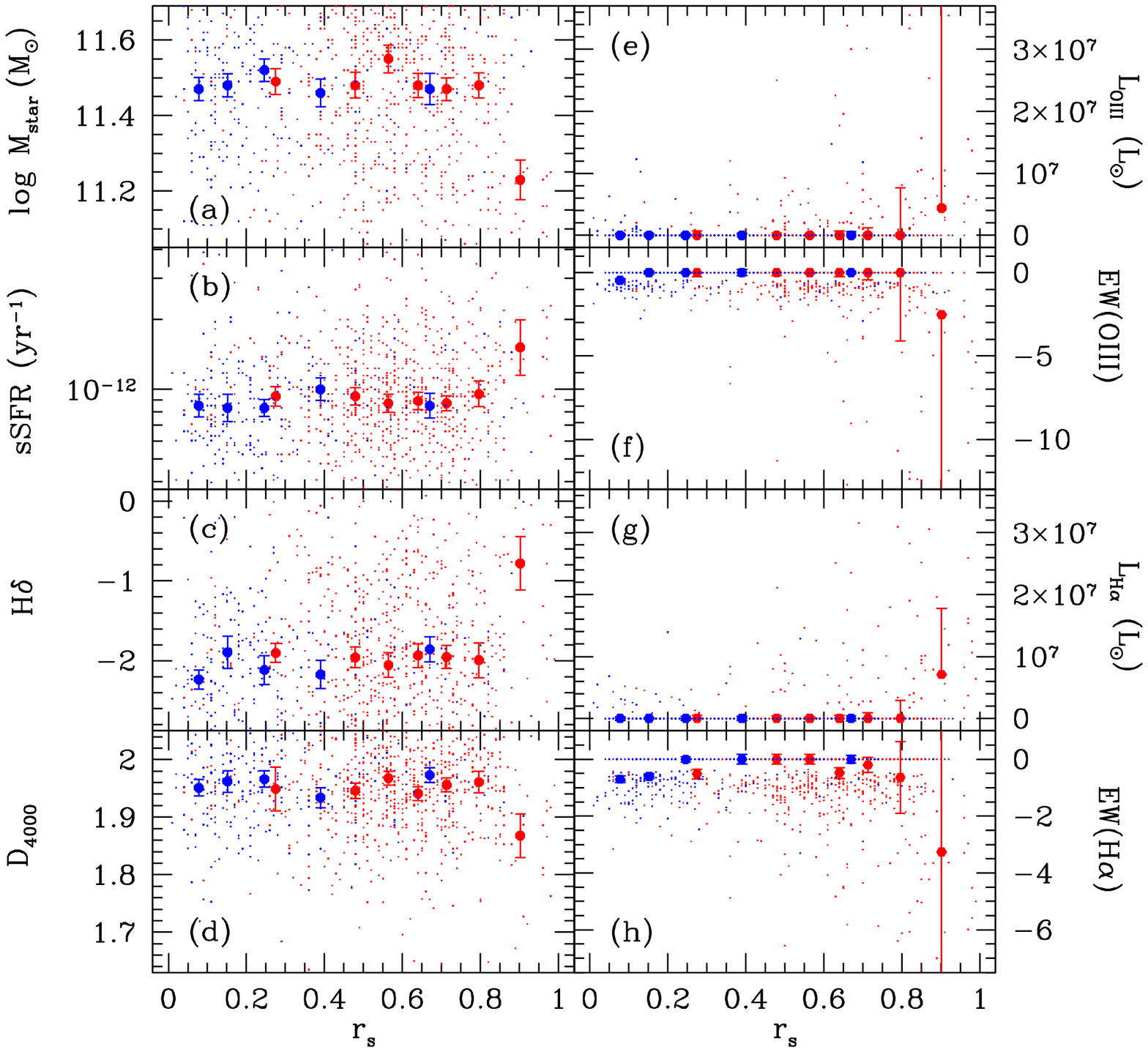}
\vspace{-10mm}
\caption{
Median values of several derived properties of the host galaxy stellar populations, as well as emission line measurements, as a function of $r_s$, for classes $a$ (red) and $b$ (blue). 
Panels (a) to (h) show the stellar mass, specific star formation rate, H$\delta$ index, 4000\AA\ break strength $D_{4000}$, [\ion{O}{3}]$\lambda5007$ line luminosity (no extinction correction applied), [\ion{O}{3}] equivalent width (EW), H$\alpha$ line luminosity, and H$\alpha$ EW, respectively. 
The errorbars show the uncertainty of the median values. 
Small points are individual RGs.
}
\label{fig:derived}
\end{figure*}

Four derived properties of the stellar populations (left panels), as well as emission line measurements (right panels), are shown in Fig.~\ref{fig:derived}. All these quantities are taken from the MPA/JHU-VAGC.
The stellar mass $M_{\rm star}$ is derived from broadband $ugriz$ photometry (see \citealt{salim07} for details). The specific star formation rate (sSFR) is derived using the method presented in \citet{brinchmann04}.
The H$\delta$ and $D_{4000}$ indices are measures of the stellar age and star formation history \citep[e.g.,][]{kauffmann03b}.
As in Fig.~\ref{fig:gal}, the $a_{0.9}$ and $a_{<0.8}$ subsets show quite different behavior. While the $a_{<0.8}$ RGs 
have roughly the same median stellar masses, sSFR, and stellar ages, and 
do not have strong emission lines,
the majority of $a_{0.9}$ objects are quite active, but less massive, and have 
higher sSFR and younger stellar age.
For example, \citet{kauffmann03} suggested the use of [\ion{O}{3}]$\lambda5007$ luminosity as an indicator of AGN activity; 26\% of the $a_{0.9}$ RGs can be regarded as strong AGNs (e.g., $\log L_{\rm OIII}>7$). We do not apply any dust-extinction correction to the line luminosities, as (1) dust content is expected to be low in the early type galaxies of our sample, and (2) any such correction (e.g., that based on the Balmer line ratios) is itself uncertain.

\begin{figure*}
\epsscale{0.75}
\plotone{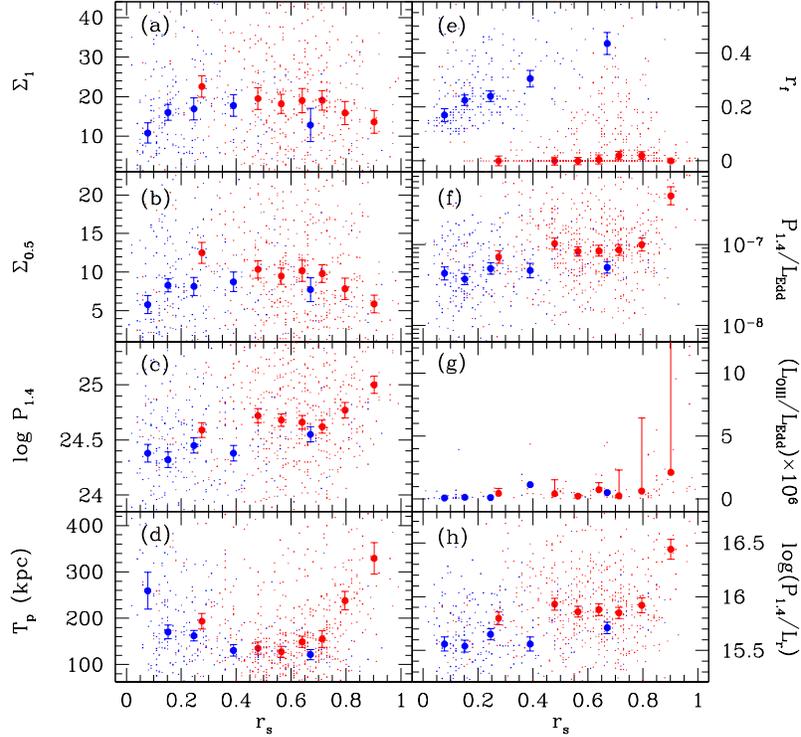}
\vspace{-10mm}
\caption{
Median values of environmental and radio properties of the host galaxies as a function of $r_s$, for classes $a$ (red) and $b$ (blue). 
Panels (a) to (h) show the excess number of neighbors over the mean background within 1 Mpc and 0.5 Mpc ($\Sigma_1$ and $\Sigma_{0.5}$), the 1.4 GHz radio power, the linear physical size of the radio sources, the central-to-total radio flux ratio, the radio and [\ion{O}{3}]$\lambda5007$ Eddington ratios, and the radio-to-optical luminosity ratio, respectively.
The errorbars show the uncertainty of the median values. 
Small points are individual RGs.
}
\label{fig:rad}
\end{figure*}

In each panel of Fig.~\ref{fig:manymp} we show as crosses the RGs whose spectra show signatures of Seyfert-type activity, and as squares those with LINER-like spectra.
We follow \citet{kauffmann03} to classify the spectra according to the [\ion{O}{3}]$\lambda5007$/H$\beta$ and [\ion{N}{2}]$\lambda6583$/H$\alpha$ line ratios on the BPT diagram \citep{baldwin81,veilleux87}.  The class $a$ sources with higher values of $r_s$ contain larger fractions of active nuclei. For example, $6/21\approx 29\%$ of the $r_s\ge 0.9$ sources have a Seyfert-like spectrum\footnote{Although there are 24 RGs with $r_s\ge 0.9$ (Fig.~\ref{fig:manymp}), only 21 have spectra of good enough quality for line measurements.}, and $2/21\approx 10\%$ can be classified as LINERs.
In Table~\ref{tab:agn} we record the fraction of RGs with Seyfert- and LINER-like nuclei as a function of $r_s$. Note that the $r_s$ binning in the Table is the same as that in Figures~\ref{fig:gal}--\ref{fig:rad}, chosen such that each bin contains roughly equal number of RGs (except for the highest $r_s$ values), and is thus different from that shown in Fig.~\ref{fig:manymp}.
While the fraction of class $a$ RGs with LINER-like spectra is roughly the same for all $r_s$ bins (at about few percent), the fraction of Seyfert-like spectrum is a strong function of $r_s$. About 20\% of the objects in the $a_{0.9}$ subset have strong emission lines that are characteristic of active nuclei.

As an indicator of the accretion rate onto the SMBH, we show in Fig.~\ref{fig:rad} (panel g) the [\ion{O}{3}] line Eddington ratio, which is the [\ion{O}{3}]$\lambda5007$ luminosity divided by the Eddington luminosity $L_{\rm Edd}$ (which is estimated using the $M_{\rm BH}$--$\sigma$ relation). 
The median accretion rate is quite close to zero for most of the $r_s$ bins, but is higher for the $a_{0.9}$ systems.
In panel (f), we show the analogous radio Eddington ratio ($P_{1.4}/L_{\rm Edd}$) at 1.4 GHz, which exhibits the same trend as $L_{\rm OIII}/L_{\rm Edd}$. Note that we do not integrate over the radio spectrum (e.g., from 0.1 to 10 GHz) to obtain the total radio power; but as we are mainly interested in the trend with $r_s$, this should not be a problem under the assumption that the spectral shape is not a function of $r_s$. Assuming a mean spectral index of $\alpha=-0.8$ for all RGs and integrating over the frequency range $\nu=0.1-10$ GHz, we would need to scale the 1.4 GHz Eddington ratio by a factor of 74 to obtain the radio Eddington ratio.

The other panels in Fig.~\ref{fig:rad} are related to the environments and other radio properties of the RGs. 
As a simple, statistical measure of the number of neighbors, for each RG we count the number of luminous galaxies ($M^*-3.5\le M_r^{0.1}\le M^*+1.5$) within 1 Mpc and 0.5 Mpc in the SDSS photometric catalog, assuming they are at the redshift of the RG, and subtract the expected number of galaxies (from the global galaxy counts) in the same apparent magnitude range. 
These neighbor counts are denoted as $\Sigma_1$ and $\Sigma_{0.5}$, respectively, and are shown in panels (a) and (b).
These two scales are chosen to reflect the scales of the host group/cluster and of any possible local structure within the group/cluster.
On average, class $a$ RGs live in dense environments (i.e., in excess with respect to the global background).
The median number of neighbors decreases with increasing $r_s$.

Panel (c) shows the mean 1.4 GHz radio power as a function of $r_s$.  Here objects in the $a_{0.9}$ subset have the highest luminosity, while $a_{<0.8}$ RGs have very similar median radio power.
Panel (d) shows the distribution of physical sizes of the RGs. Not surprisingly, the RGs with highest $r_s$ are also largest (median size exceeding 300 kpc), since it is mainly the lobes that contribute to the low-frequency radio fluxes.

Panel (e) is another measure of the radio source morphology: the central-to-total flux ratio $r_f$. We estimate the central flux by summing fluxes from all components within $0.15T$ from the center. For class $a$, the $r_f$ ratio is on average small ($<0.1$). Note that this must be partially a selection effect, for if a strong source is present at the center,
it will likely be the HSB spot and
the RG would be put in other classes ($b-e$).

Finally, panel (h) shows the logarithm of the radio-to-optical luminosity ratio (calculated simply  as $\log P_{1.4}+0.4 M_r^{0.1}$), which 
is also almost constant at $r_s\lesssim 0.8$, and rises sharply at highest $r_s$ bin; this  enhancement for $a_{0.9}$ is due to the combined effect of it having the 
lowest median 
optical luminosity and highest median radio power among the class $a$ objects.

All the results from Figures~\ref{fig:gal}--\ref{fig:rad} indicate that there appear to be two populations of RGs among class $a$: galaxies in the $a_{0.9}$ subset have lowest stellar mass, but have the highest star formation rate, AGN activity, and radio power, and live in relatively poorer environments, while 
the $a_{<0.8}$ RGs, especially those with $r_s\approx 0.4-0.8$, share very similar properties, such as the luminosity, size, dynamical mass, density, indicators of recent star formation history (e.g., $D_{4000}$ and H$\delta$ indices), and neighbor counts within 0.5 and 1 Mpc. 
We should emphasize that the division in $r_s$ of the two populations is not sharp; we choose to distinguish the two subsets at $r_s=0.8$ mainly for simplicity.

Can we identify these two populations with the two FR types? 
The general properties of these two populations do conform roughly to the characteristics of the two FR types found in the literature (\S\ref{sec:intro}). 
Before making a direct correspondence (see \S\ref{sec:threetypes}), however, it is important to investigate where the class $b$ objects fit in the context of the FR I/II dichotomy.

%%%%%%%%%%%%%%%%%%%%%%%
\subsection{Class $b$}
\label{sec:classb}

The $r_s$ binning in Figures \ref{fig:gal}--\ref{fig:rad} for the class $b$ (blue points) is the same as that shown in Fig.~\ref{fig:manymp} (right panels); the numbers of RGs in the five bins are (from lowest to highest): 57, 78, 68, 44, and 42.
The median redshift slightly increases with $r_s$ (Fig.~\ref{fig:gal}, panel j; see Appendix \ref{sec:first}).

Broadly speaking,
the class $b$ RGs are typical early type galaxies in their optical properties (based on e.g., the color, concentration, and Sersic index).
One gets an impression 
from Figures \ref{fig:gal}--\ref{fig:rad} that there is not much variation with $r_s$ for galaxies in class $b$. 
For example, $r_{\rm eff}$, $M_{\rm dyn}$, $\rho$, $\mu$, and 1.4 GHz Eddington ratio all have median values that do not vary much with $r_s$.
The most notable systematic variations are that $(g-r)^{0.1}$ color becomes redder as $r_s$ increases,
a positive correlation between $r_f$ and $r_s$, and a negative correlation between the total size of the radio source and $r_s$.

There is considerable overlap in the distributions of various physical properties between the $a_{<0.8}$ and class $b$ RGs. However,
the class $b$ objects have slightly higher median $\sigma$, $c$, $\rho$, $\mu$, smaller $r_{\rm eff}$, lower emission line strength, neighbor counts, radio power and Eddington ratio, and radio-to-optical luminosity ratio. That is, they are more compact and are more quiescent in terms of their nuclear and radio activity.  
Of the 597 class $a$ objects for which classification of spectral properties using the BPT diagram is possible, 37 ($=6.2\%$) are Seyferts or LINERs. For class $b$, the fraction is only $4/256=1.6\%$.

In Table~\ref{tab:maintable} we list the basic properties of the RGs in our sample.

%%%%%%%%%%%%%%%%%%%%%%%
\subsection{Environments of Extended Radio Galaxies}
\label{sec:environment}

From Fig.~\ref{fig:rad} we see that
%In particular, 
the $a_{<0.8}$ subset has the highest neighbor counts (median $\Sigma_{0.5} \approx 9.9\pm 0.6$), while those of class $b$ and $a_{0.9}$ are close ($\Sigma_{0.5}\approx 7.4\pm 0.6$ and $6.7\pm 1.1$, respectively). For reference, for the most luminous galaxies ($M_r^{0.1}=-22.77$ to $-24.27$, irrespective of radio properties), $\Sigma_{0.5}\approx 10.8\pm 1.0$.

To better understand the environments of these RGs, we further examine their association with galaxy clusters, and calculate their clustering properties.
First, we cross-match our RG sample with the clusters found by the maxBCG algorithm \citep{koester07}. 704 ($=75\%$) of our class $a$ and $b$ RGs lie in the footprint and the redshift range ($z=0.1-0.3$) of the maxBCG survey.
We focus on clusters more massive than $\approx 10^{14} M_\odot$, above which the cluster catalog is about $\sim 80\%$ complete. The cluster mass and the virial radius are estimated using the weak lensing-calibrated mass--observable scaling relation from \citet{reyes08}. Only 13\%, 24\%, and 14\% of the RGs in the $a_{0.9}$, $a_{<0.8}$, and $b$ subsets are associated (i.e., within one virial radius) with these optically-selected clusters\footnote{The class $a$ and $b$ RGs only account for $<10\%$ of our full RG sample.
The narrow-angle tail objects, whose morphology is believed to be due to the relative motion of the galaxies and the intracluster medium, together with the compact sources (which account for the majority of the sources), are excluded in the current sample.
The fraction of all RGs that are cluster members is therefore higher than we have estimated here.}. 
However, $>80\%$ of those RGs that are in clusters lie within 20\% of the virial radius from the cluster center, consistent with previous findings that RGs are centrally concentrated in clusters \citep[e.g.,][]{ledlow95,lin07}.

\begin{figure}
%\epsscale{0.9}
\plotone{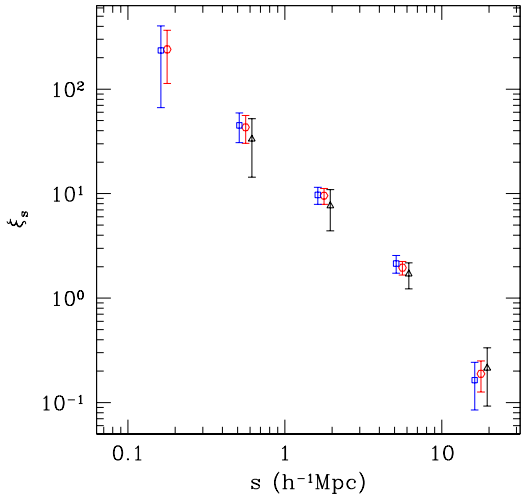}
%\vspace{-1mm}
\caption{
Redshift space RL-RQ cross correlation functions 
for the $a_{0.9}$ (black triangles), $a_{<0.8}$ (red circles), and $b$ (blue squares) subsamples. 
These are calculated using galaxies that form a volume-limited sample ($z\le 0.16$, $M_r^{0.1}\le -21.77$; for RGs an additional criterion is $\log P\ge 23.31$).
At scales larger than $\approx 200 h^{-1}\,$kpc, both $a_{<0.8}$ and class $b$ have similar clustering strength, suggesting that they are hosted by halos of similar mass. 
The $a_{0.9}$ subset might be hosted by halos of lower mass, although a larger sample is needed to test this possibility.
Some $a_{0.9}$ RGs have similar optical properties and likely also similar environment as the other two subsets (see \S\ref{sec:threetypes}).
}
\label{fig:xi}
\end{figure}

\begin{figure*}
\epsscale{0.75}
\plotone{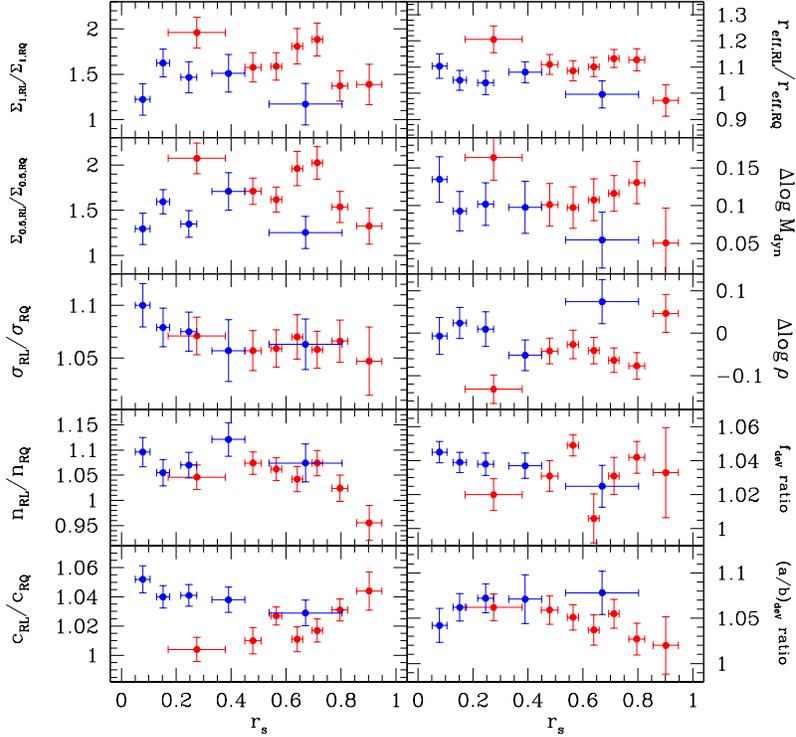}
\vspace{-12mm}
\caption{
Differences between RL and RQ galaxies. Red and blue points denote classes $a$ and $b$, respectively. For every RG we find up to ten RQ galaxies that have very similar $M_g^{0.1}$, $M_r^{0.1}$, and $M_i^{0.1}$ magnitudes and redshift; after calculating the mean value of a given physical quantity from all RGs in a given $r_s$ bin, we then take the ratio (or the difference if the quantity is logarithmic) between the mean value of the RGs and that obtained from all the RQ galaxies that are matched to the RGs in question. The left panels, from top to bottom, show the ratio in neighbor counts within 1 Mpc and 0.5 Mpc, the velocity dispersion, Sersic index, and concentration. The right panels, from top to bottom, show the ratio/difference in effective radius, log dynamical mass, log mass density, contribution of the de Vaucouleurs profile to the SB profile (normalized to unity), and the axis ratio. The errorbars denote the uncertainties in the mean value.
}
\label{fig:rlrq1}
\end{figure*}

The RGs are mainly hosted in dark matter halos more massive than $\approx 2\times 10^{13} M_\odot$ \citep{mandelbaum09},
so the small fraction of RGs associated with maxBCG clusters suggests
that the majority of our RGs must be associated with clusters or groups of mass $\approx 2-10 \times 10^{13} M_\odot$, a range in which the maxBCG catalog is highly incomplete \citep{koester07}.  
We then resort to the clustering properties of the RGs, which provides some insight into the 
relative mass scales of halos that host these subsets of RGs.
We calculate the cross correlation functions between the RG subsets and the general galaxy population, constructed as a volume-limited sample of 73,202 galaxies from the NYU-VAGC DR6 ($z\le 0.16$, $M_r^{0.1}\le -21.77$).
We select the RGs to satisfy the same redshift and magnitude cuts as well as a lower limit in radio power $\log P_{1.4}\ge 23.31$, which results in 41, 211, and 123 objects for $a_{0.9}$, $a_{<0.8}$, and $b$ subsets, respectively. The redshift space cross correlation functions are shown in Fig.~\ref{fig:xi}, for $a_{0.9}$ (black triangles), $a_{<0.8}$ (red circles), and class $b$ (blue squares). 
Although all three subsets have similar clustering strengths
at scales $\gtrsim 200 h^{-1}\,$kpc, 
there is a slight hint of lower clustering amplitude for the $a_{0.9}$ objects; if this is confirmed with larger RG samples, this implies the host halos of the $a_{0.9}$ subset are on average less massive than the hosts of the other subsets.

Combining these results, we see that the distributions of host halo mass for the $a_{<0.8}$ and class $b$ subsamples are similar at group mass scale, but that of the former must have a higher tail towards clusters ($>10^{14} M_\odot$). For RGs in groups, 
the $a_{<0.8}$ objects must be more centrally concentrated than the class $b$.

%%%%%%%%%%%%%%%%%%%%%%%%%%%%%%%%%%%%%%%%%%%%%%
\subsection{Radio-Loud \lowercase{vs} Radio-Quiet}
\label{sec:rlvsrq}

So far we have made comparisons among different subsets of RGs. It is important to place them in the context of general massive galaxy populations.
Furthermore, some of the correlations between the physical properties and $r_s$ may be due to other fundamental correlations of the early type galaxies that have no physical connection to the radio source. For example, given that optical luminosity and effective radius are tightly correlated in elliptical galaxies \citep[e.g.,][]{shen03}, an anti-correlation between $M_r$ and $r_s$ would imply a similar anti-correlation between $r_{\rm eff}$ and $r_s$ (Fig.~\ref{fig:gal}).
To take out such an effect, for every RG we find up to ten RQ galaxies that have very similar redshift and absolute magnitudes in the $g^{0.1}$, $r^{0.1}$, and $i^{0.1}$ bands. 
For all RGs in a given $r_s$ bin, we calculate the average value of the physical property in question from all the matched RQ galaxies and subtract that value off from the mean obtained for the RGs (or take the ratio of the two, depending on the nature of the properties).
Note that some AGNs (Seyferts and LINERs) may be included in the RQ sample, as we do not distinguish truly quiescent galaxies from those that can be regarded as AGNs based on the optical emission line diagnostics.

\begin{figure*}
\epsscale{0.75}
\plotone{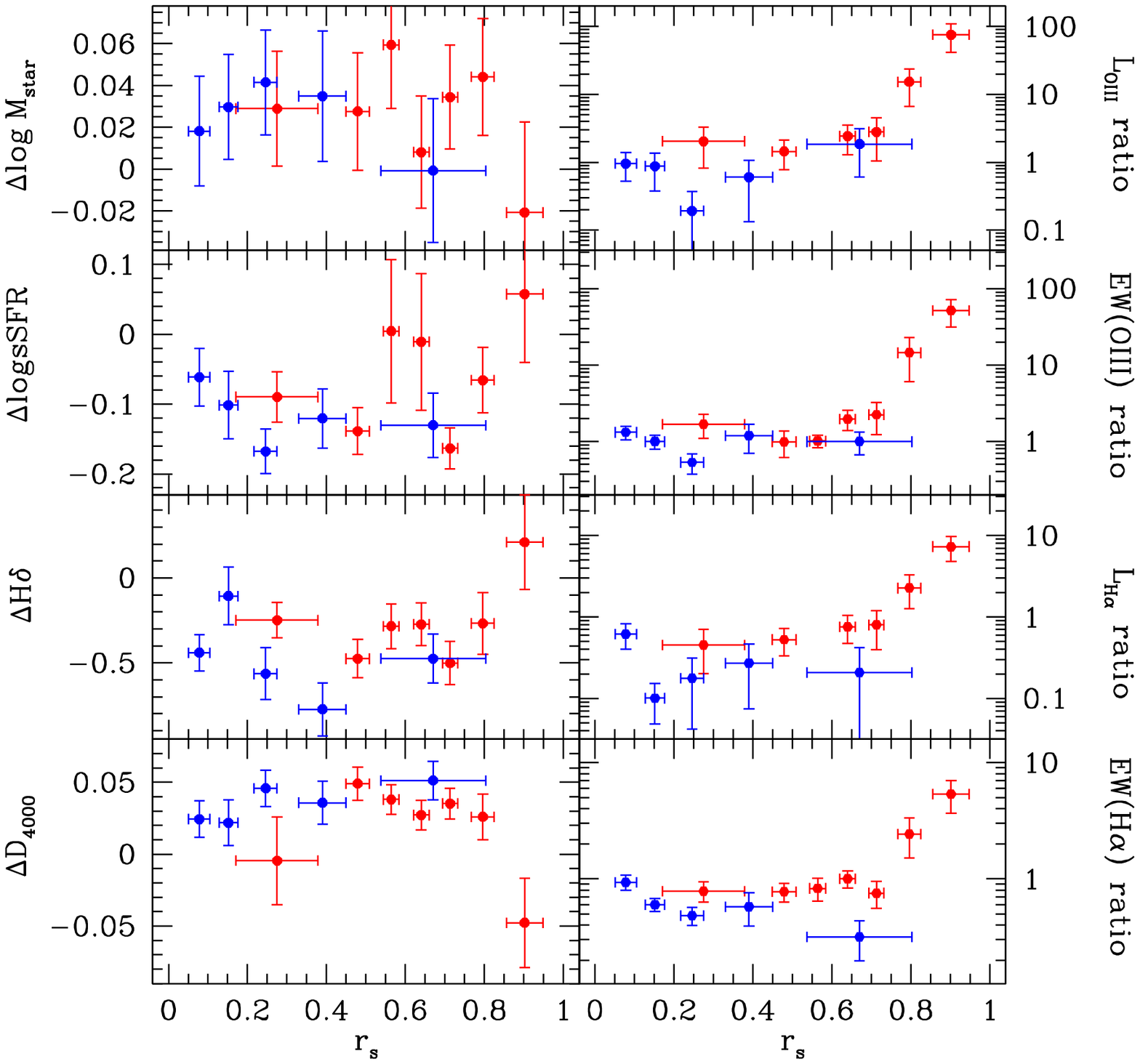}
\vspace{-12mm}
\caption{
Differences between RL and RQ galaxies. Red and blue points denote classes $a$ and $b$, respectively. The left panels (from top to bottom) show the differences (RL minus RQ) in stellar mass, specific star formation rate, H$\delta$ index, and 4000\AA\ break strength. The right panels (from top to bottom) show the line ratios (RL divided by RQ) for [\ion{O}{3}] and H$\alpha$. See caption of Fig.~\ref{fig:rlrq1} for details of the RL-RQ matching. The errorbars denote the uncertainties in the mean value.
}
\label{fig:rlrq2}
\end{figure*}

Figures \ref{fig:rlrq1} and \ref{fig:rlrq2} 
show the results as a function of $r_s$.
The comparisons presented in these two Figures are between each class and its RQ counterparts, not between the two classes.
For most of the properties shown in Fig.~\ref{fig:rlrq1} (environments and global galaxy properties), RL galaxies have higher mean values than do the RQ ones (e.g., more neighbors, higher mass and velocity dispersion, larger effective radius).
The only exception is the central mass density/stellar surface density, for which the $a_{<0.8}$ subset appears to be less dense than its RQ counterparts, mainly due to their larger scale lengths.
Fig.~\ref{fig:rlrq2} shows comparisons for some derived quantities of the stellar population and the line measurements. The $a_{<0.8}$ and class $b$ RGs on average have slightly higher stellar mass, and older stellar age ($\Delta{\rm H}\delta={\rm H}\delta_{\rm RL}-{\rm H}\delta_{\rm RQ}<0$), than their respective RQ counterparts.

Among the class $a$ objects,
galaxies in the $a_{0.9}$ subset are closest to their RQ matches in terms of neighbor counts, $r_{\rm eff}$, $M_{\rm dyn}$, and $\mu$, but show dramatically stronger emission lines and sSFR.
In contrast, emission lines are weaker for class $b$ RGs than their RQ counterparts.
Class $b$ objects also 
live in environments that are closer to their RQ counterparts than those of class $a$, and their sizes ($r_{\rm eff}$) are also more similar.

\begin{figure*}
%\begin{figure}
\epsscale{0.74}
%\epsscale{1.37}
\plotone{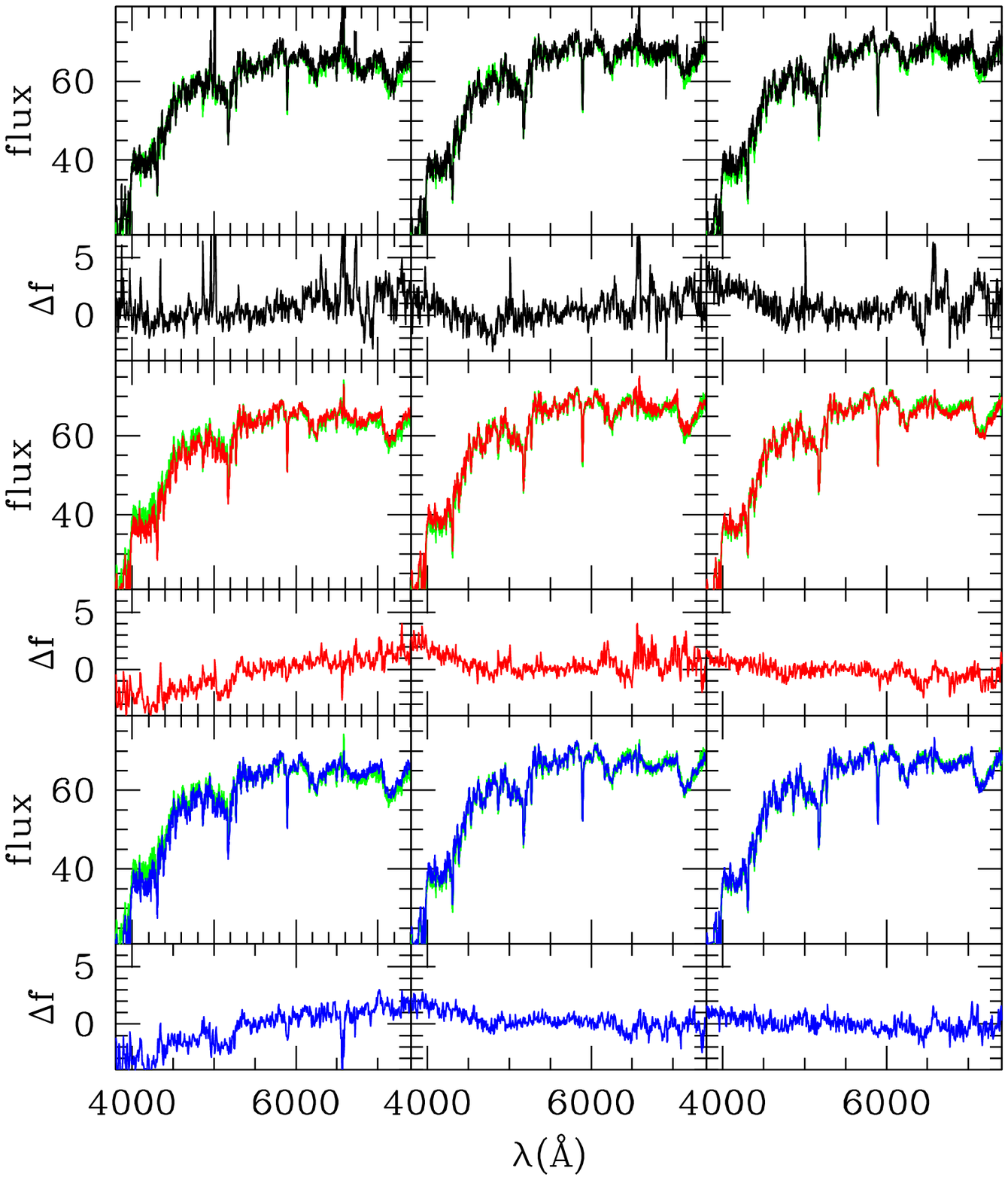}
%\hspace{-5mm}
\vspace{-9mm}
\caption{
Mean SDSS spectra of RGs of different morphologies and different SMBH masses. From left to right, the columns are of increasing ranges of $M_{\rm BH}$ (1--$3\times 10^8 M_\odot$, 3--$5\times 10^8 M_\odot$, 5--$8.5\times 10^8 M_\odot$). Within each column, the top two panels show the mean spectrum of $a_{0.9}$ objects (black curve) and its difference from the RQ galaxies (green curve, repeated in other panels in a column) with similar $M_{\rm BH}$, $\mu$, and $M_{\rm dyn}$. The two panels in the middle are for $a_{<0.8}$ RGs (red curve); the bottom two are for class $b$ (blue curve). Each pair of RL and RQ spectra is normalized at 5500--5600\,\AA. The mean spectra are smoothed by a 5\AA\ boxcar.
}
\label{fig:bhbins}
\end{figure*}
%\end{figure}

Let us examine the spectral properties of the RGs in more detail.
In Table~\ref{tab:agn}, in parentheses, we list the fraction of the matched RQ galaxies that show Seyfert- and LINER-like spectrum. This is useful for evaluating whether the active nucleus fraction is elevated in a given $r_s$ bin. For example, for class $a$ with $r_s\ge 0.85$, 19.6\% of RGs have Seyfert nuclei, while only 1.5\% of the matched RQ galaxies exhibit the same level of activity. In the same $r_s$ bin, 4.3\% (4.9\%) of the RL (RQ) galaxies have LINER nuclei.
The LINER fraction for class $a$ is roughly independent of $r_s$, and is fully consistent with what is found in the RQ populations, but the Seyfert activity is enhanced by about a factor of ten in RGs with $r_s\gtrsim 0.6$ (see also \citealt{ivezic02}).
There is some suggestion that the nuclei of class $b$ galaxies are actually more {\it quiescent} than the matched RQ galaxies. For example, there are 2659 RQ galaxies matched to the class $b$ subsample, out of which 84 are LINERs. We thus expect to find about 8 LINERs in the 256 class $b$ RGs, but only detect 3, which is inconsistent with the Poisson expectation at 99\% level \citep{gehrels86}.

The differences in the optical spectral features are further illustrated in Fig.~\ref{fig:bhbins}, which shows stacked SDSS spectra of various subsamples of RGs.
To make a fair comparison among the three subsets ($a_{0.9}$, $a_{<0.8}$, $b$), as well as between RQ and RL galaxies, we select galaxies in three SMBH mass (stellar velocity dispersion) bins, and limit the ranges of dynamical mass and stellar mass surface density to be within 50\% of the $a_{0.9}$ locus\footnote{For computing the mean spectra, the RQ galaxies are selected to match $\sigma$, $M_{\rm dyn}$, and $\mu$ of the RGs, without any restriction on redshift and restframe $g^{0.1}$, $r^{0.1}$, and $i^{0.1}$ magnitudes, as was done earlier in this section.}.
These properties are chosen to select RGs of similar central engine, fuel supply, and structure.
There are (19,46,23) RGs in the lowest mass bin for the $a_{0.9}$, $a_{<0.8}$, and $b$ subsets, respectively. In the intermediate and high mass bins, the numbers of RGs are (11,99,63) and (11,71,40).
In the Figure,
the three columns correspond to the three mass bins (in increasing mass order from left to right); three pairs of panels are shown in each column (from top to bottom: $a_{0.9}$, $a_{<0.8}$, $b$). The pair consists of the mean 
spectra of an RG subsample and the RQ galaxies of similar $M_{\rm BH}$, $M_{\rm dyn}$, and $\mu$ (upper panel) and the differences between the RL and RQ spectra (lower panel).
For each $M_{\rm BH}$ bin, the RQ galaxy mean spectrum is an average over 80 randomly selected galaxies, which are chosen 
irrespective of their spectral properties, and thus may contain some AGNs (Seyferts and LINERs).

A few points are worth noting
from the RL$-$RQ spectral difference panels. 
(1) Only $a_{0.9}$ RGs have statistically significantly stronger emission lines than their RQ counterparts. 
(2) The emission line (particularly 
[\ion{O}{3}]$\lambda5007$ and
H$\alpha$)
strength decreases as $M_{\rm BH}$ increases for $a_{0.9}$ objects.
(3) In the lowest mass bin, $a_{<0.8}$ and class $b$ RGs are redder than their RQ counterparts (based on the difference spectrum); this is not seen in the other bins.\footnote{Recall that in Figures \ref{fig:rlrq1} and \ref{fig:rlrq2} the optical colors [$(g-r)^{0.1}$, $(r-i)^{0.1}$] are used to match the RL and RQ galaxies, and thus we do not examine color differences.}
In addition, the H$\alpha$ line is 
much weaker than that in the RQ galaxies.
(4) In each mass bin, $a_{<0.8}$ and class $b$ have, to first order, similar difference spectra, indicating their spectra are close to each other.

%%%%%%%%%%%%%%%%%%%%%%%%%%%%%%%%%%%%%%%%%%%%%%
\section{Discussion}
\label{sec:disc}

In this section we build upon the observational results presented in the previous two sections to investigate some intriguing questions related to the physical nature of RGs: {\it What is the correspondence between the FR types and the classes $a$ and $b$?} {\it Do class $b$ or $a_{<0.8}$ objects represent evolutionary sequences?}  {\it What is the physical origin of various morphologies?}

%%%%%%%%%%%%%%%%%%%%%%%
\subsection{Three Types of RGs?}
\label{sec:threetypes}

46 out of 85 $a_{0.9}$ objects have no detectable [\ion{O}{3}]$\lambda5007$ emission line (i.e., the signal-to-noise ratio of the line is less than three). 
How do they differ from those objects with emission lines?
About $75\%$ of $a_{0.9}$ RGs with [\ion{O}{3}] line luminosity $>10^6 L_\odot$ (roughly corresponding to [\ion{O}{3}] Eddington ratio of $>10^{-7}$) show clear hot spots at the edge of the lobes (giving the impression of a bullet shot into a tenuous medium), while about $2/3$ of the $a_{0.9}$ objects without emission lines have HSB spots that show less contrast with the lobes, or have lobes that are not well-aligned. 
%
%It seems plausible that the mechanism that creates the emission lines is physically related to the generation of hot spots.
It is plausible that the mechanism that creates the emission lines is physically related to the process responsible for the generation of hot spots.
%%% NOTE!!!

The distributions of many physical properties for the $a_{0.9}$ RGs with and without emission lines are often offset from each other (albeit with substantial overlap).
It is the $a_{0.9}$ RGs with emission lines (hereafter $a_{\rm 0.9,em}$ objects; Table~\ref{tab:subclasses}) that make this subset stand out from the rest of RGs.
%
%It is the $a_{0.9}$ RGs with emission lines (hereafter $a_{\rm 0.9,em}$ objects; Table~\ref{tab:subclasses}) that make this subset stand out from the rest of RGs;
%for many of the physical properties, their distribution is offset from that of the quiescent $a_{0.9}$ objects (although there is still substantial overlap in the distributions).
On the other hand, the $a_{0.9}$ objects
without emission lines share very similar properties with the $a_{0.8}$ subset; the median values of most physical properties are within $1\sigma$ of each other. We may regard them as the same population as the $a_{<0.8}$ subsample,
and refer to the combined population, which accounts for the majority of class $a$ objects, as $a_{\rm maj}$ (see Table~\ref{tab:subclasses}).

We emphasize that the $a_{\rm 0.9,em}$ objects are still massive galaxies ($M_r\lesssim M^*$), and that the distinction of this population from the rest of RGs is far less dramatic than the red--blue galaxy bimodality of the general galaxy population (e.g., \citealt{baldry04}).

%The $a_{0.9}$ objects
%without emission lines share very similar properties with the rest of the $a_{0.8}$ subset; the median values of most physical properties are within $1\sigma$ of each other. We may regard them as the same population as the $a_{<0.8}$ subsample,
%and refer to the combined population, which accounts for the majority of class $a$ objects, as $a_{\rm maj}$ (see Table~\ref{tab:subclasses}).
%
%It is thus the $a_{0.9}$ RGs with emission lines (hereafter $a_{\rm 0.9,em}$ objects; Table~\ref{tab:subclasses}) that make this subset stand out from the rest of RGs.
%For many of the physical properties, their distribution is offset from that of the quiescent $a_{0.9}$ objects (although there is still substantial overlap in the distributions).
%
%We emphasize that the $a_{\rm 0.9,em}$ objects are still massive galaxies ($M_r\lesssim M^*$), and that the distinction of this population from the rest of RGs is far less dramatic than the red--blue galaxy bimodality of the general galaxy population (e.g., \citealt{baldry04}).

Since our proposed classification scheme combines both radio morphology and nuclear emission line strength, 
while previous ones usually rely on one or the other of these criteria (see \S\ref{sec:intro}), a perfect correspondence between the two is not expected.
We also emphasize that our scheme is more quantitative, objective, and reproducible, than are either the FR or OL89 classifications.

Broadly speaking, we can identify the $a_{\rm 0.9,em}$ objects with HE RGs, and the rest in our sample with LE RGs, with the caveat that some of the $a_{\rm maj}$ objects do have strong emission lines (c.f.~Table~\ref{tab:agn}).
Regarding the classification scheme of OL89, the $a_{\rm 0.9,em}$ subset corresponds to the CD type (mainly because of the high occurrence of hot spots of the former), the $a_{\rm maj}$ subset is consistent with the FD type, and class $b$ coincides with the TJ type.
%
%Broadly speaking, the $a_{\rm 0.9,em}$ subset corresponds to the CD type (mainly because of the high occurrence of hot spots of the former), the rest of class $a$ (the majority, hereafter $a_{\rm maj}$; see Table~\ref{tab:subclasses}) is consistent with the FD type, and class $b$ coincides with the TJ type as defined by OL89.
%
Although the average optical properties of class $b$ and $a_{<0.8}$ (or $a_{\rm maj}$) RGs are similar, 
we regard them as distinct populations, mainly based on
the differences in their environment, radio, and nuclear activity.

By definition, the $a_{\rm 0.9,em}$ objects have $r_s>0.8$, and are thus associated with FR IIs according to the original FR definition. A direct correspondence between the FR types and our $a_{\rm maj}$ and $b$ subsets is not possible, however, given that the $r_s$ distributions for both class $b$ and $a_{\rm maj}$ objects are quite broad (e.g., Figures~\ref{fig:gal}--\ref{fig:rad}).
%
%If we loosen the definition of FR types and adopt the stereotype as mentioned in \S\ref{sec:intro}, 
%If we adopt the apparent trends discussed in \S\ref{sec:intro} between FR
%type, luminosity, and emission-line strength,
%a better ``match'' can be found between our $a_{0.9}$, $a_{<0.8}$, and $b$ scheme and the FR types.
%
%Given the similarities of class $b$ and $a_{<0.8}$ objects, and the large difference between these subsamples and the $a_{0.9}$ subset, we would identify the FR I type with the former and FR II with the latter sources. %(if we were to ignore the fact that a nonnegligible fraction of FR IIs do not have appreciable nuclear emission lines).
%
If we have to adopt a dichotomy classification scheme, as advocated by FR and followed by many others, then we may call the class $b$ plus $a_{<0.8}$ objects type I, and the $a_{0.9}$ objects type II. However, referring to these two groups as ``FR I'' and ``FR II'' would be mis-leading, as the original FR definition is solely based on $r_s$.

To summarize, based on the properties of the host galaxies and radio emission, we suggest there are three groups of RGs in our sample: $a_{\rm 0.9,em}$, $a_{\rm maj}$, and class $b$.
There is no single physical property that can be used to cleanly separate one group from the others. 
For example, %while the $a_{\rm 0.9,em}$ subset by definition has high $r_s$ values, 
the $r_s$ distribution of the $a_{\rm maj}$ group is quite broad (c.f.~Fig.~\ref{fig:sthist}), almost encompassing that of the $a_{\rm 0.9,em}$ objects at the high $r_s$ end.
{\it A simple morphological measure such as $r_s$ is thus only of limited use for classifying extended RGs.}

In addition to $r_s$, we have also explored the use of radio power $P_{1.4}$ in the classification scheme. Although the $\sim 5\%$ of the RGs with highest radio power are reasonably separated from the rest in plots like Figures~\ref{fig:gal}--\ref{fig:rad} (with abscissa replaced with $P_{1.4}$), suggesting that selecting via $P_{1.4}$ can in principle producing a subsample similar to $a_{0.9}$, we decide to stick with the morphological parameter $r_s$, as understanding the origin of differences in the radio morphology is one of the main objectives of this paper (see \S\ref{sec:origin}).

We conclude by estimating the abundances of the three morphological groups, using
the same volume-limited sample as in \S\ref{sec:environment} ($z\le 0.16$ and $M_r^{0.1}\le -21.77$; for RGs an additional requirement is $\log P_{1.4}\ge 23.31$).
The abundances relative to all galaxies are ($a_{\rm 0.9,em}, a_{\rm maj}, b$) $=$ (0.034\%,0.41\%,0.20\%). 
Among the RGs (irrespective of morphology/extendedness), the fraction of these types are 0.9\%, 11.1\%, and 5.5\%, respectively (see also Table~\ref{tab:subclasses}).

%% g in VL sample=73202
%% rg in VL sample=2686

\begin{figure}
\epsscale{0.9}
\plotone{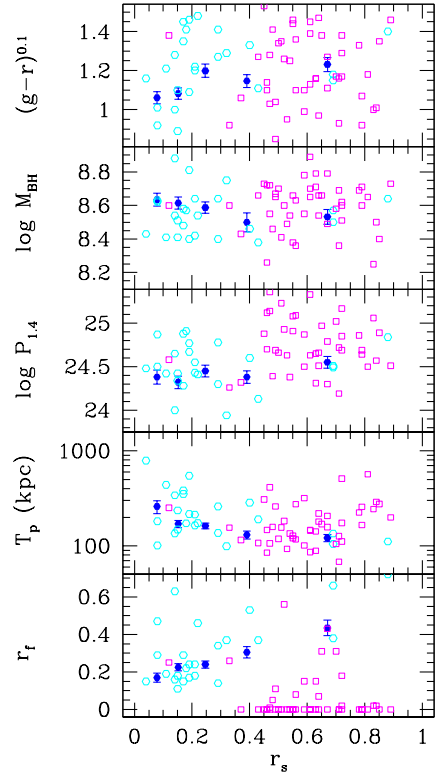}
\vspace{-4mm}
\caption{
Illustration of a possible evolutionary sequence for class $b$ RGs (cyan open circles). These objects are selected to have $M_r^{0.1}$, $M_{\rm dyn}$, $\mu$, and $\Sigma_{0.5}$
similar to the median value of class $b$ RGs in the $r_s=0.2-0.3$ bin. The blue solid points and the errorbars are the median value and its uncertainty for all class $b$ objects.
It is plausible that during the life of a class $b$ source, both $r_s$ and $r_f$ decrease with age, while $T_p$ increases, as implied by the average trends of the cyan points. The magenta squares are the $a_{<0.8}$ objects with similar mass, structure, and neighbor counts as the cyan points. 
They do not follow the trends of the cyan points, suggesting that
for this particular choice of mass and environment, 
these class $a$ objects do not represent the later phases of evolution for class $b$ objects. 
}
\label{fig:ev}
\end{figure}

%%%%%%%%%%%%%%%%%%%%%%%
\subsection{Evolutionary Sequences?}
\label{sec:evol}

The small dependence of the majority of physical properties we have examined on $r_s$ for
the class $b$ objects prompts the question: are they the same RGs viewed at different stages of evolution?
One could imagine that a young RG starts with large $r_s$ and $r_f$ (and smallest total size $T_p$); as the jets/lobes advance, $r_s$ and $r_f$ both decrease, while $T_p$ grows.
Such a trend seems to be present in Fig.~\ref{fig:rad}.
It is important to realize that, however, many such evolutionary sequences (of different combinations of the host galaxy, central engine, environments, etc) are probably simultaneously present in our sample, and therefore the median behavior of the class (as seen in Figures~\ref{fig:gal}--\ref{fig:rad}) may not reflect any one sequence.
To single out an evolutionary sequence, one should therefore only consider RGs of very similar properties (at least those properties that will not change over radio lobe time scales), such as mass, structure, and neighbor counts.

We test this idea in Fig.~\ref{fig:ev}. 
The cyan open points are a subset of galaxies in class $b$ selected to have $M_r^{0.1}$, $M_{\rm dyn}$, $\mu$, and $\Sigma_{0.5}$ similar to the median value of the $r_s=0.2-0.3$ bin.  
If they can be regarded as an evolutionary sequence parameterized by decreasing $r_s$, we would expect $r_f$ to decrease while $T_p$ increases, which is in rough agreement with the observed trends (of the open points),
although we caution the trends may be somewhat driven by the few objects at larger $r_s$.\footnote{Another caveat is that the surface brightness of the class $b$ objects may decrease rapidly outwards, producing correlations between $r_s$, $T_p$ and $r_f$ similar to that due to an evolutionary sequence for sources close to the detection limit. We thank Philip Best for pointing this out. A larger sample, with careful selection criteria, would be needed to assess the contamination due to this effect.}
Even though their size becomes bigger, their radio power stays about the same, probably due to the fast dissipation of energy in the jets; the outer regions do not contribute much to the luminosity.

We saw in \S\ref{sec:classb} that there is significant overlap in the distributions of various properties for class $b$ and $a_{<0.8}$ RGs. 
It is possible that some of the $a_{<0.8}$ RGs represent the later phases of evolution of class $b$ objects.
In Fig.~\ref{fig:ev} we show as magenta open squares the $a_{<0.8}$ objects with the same ranges of $M_r^{0.1}$, $M_{\rm dyn}$, $\mu$, and $\Sigma_{0.5}$ as the class $b$ objects (cyan points).
The most notable trend with $r_s$ is $T_p$.
If these subsets of the two classes were related, $r_s$ needs to increase as the sources age. 
However, the higher typical radio power of the $a_{<0.8}$ objects make such an evolutionary scenario implausible.

%%%%%%%%%%%%%%%%%%%%%%%
\subsection{The $P$--$M$ Plane Revisited}
\label{sec:mpagain}

In \S\ref{sec:owen} we noted that RGs in our sample 
are not separated
into two groups in the $P$--$M$ plane via a simple division in $r_s$. With the correspondence between our three subsets and the three morphological groups identified by Owen and co-workers (\S\ref{sec:threetypes}), could we better reconcile their results with ours?

\begin{figure}
\epsscale{1.15}
%\plotone{fig_oer_jun25.eps}
\plotone{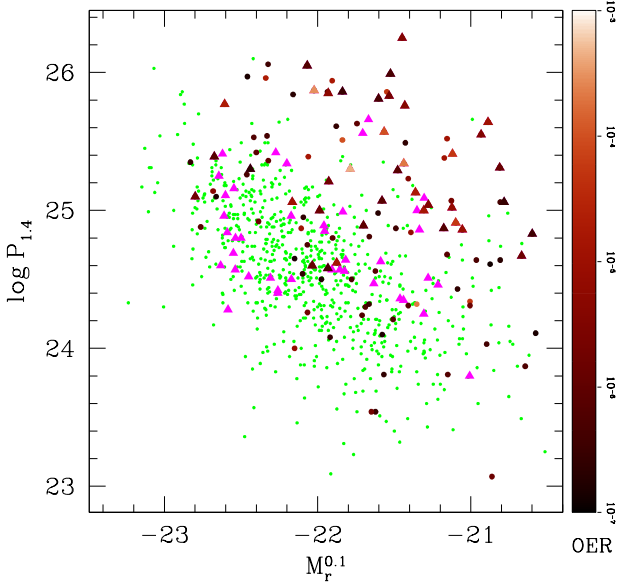}
\vspace{-3mm}
\caption{
Distribution of extended RGs in the $P$--$M$ plane. RGs without appreciable [\ion{O}{3}]$\lambda5007$ emission are shown as small green points.  Those that do are represented as larger points whose color reflects the [\ion{O}{3}]$\lambda5007$ Eddington ratio (OER), as indicated on the color bar on the right hand side. $a_{0.9}$ objects without and with OER measurements are shown as triangles in magenta and in orange of various degrees of intensity, respectively.
}
\label{fig:oermp}
\end{figure}

In Fig.~\ref{fig:oermp},
the small green dots show all the extended RGs in our sample, and triangles are the $a_{0.9}$ subset. 
The color bar in this Figure represents the [\ion{O}{3}]$\lambda5007$ line Eddington ratio (OER).
There is a better, although still not complete, separation in the $P$--$M$ plane of the $a_{\rm 0.9,em}$ RGs 
with relatively higher level of nuclear activity (e.g., OER $>10^{-6}$, roughly corresponding to $L_{\rm OIII}>10^7 L_\odot$) from the rest of the population.
%
%A slanted line on this plane {\it roughly} separates the RGs with high and low OER (with a division at $\approx 10^{-6}$) independent of radio morphology.

The significant overlaps among the different classes of RGs in the parameters we have surveyed (e.g., $P_{1.4}$, $M_r^{0.1}$, mass, structure, environment; Figures~\ref{fig:gal}--\ref{fig:rad}) imply that no simple combination of non-radio observables can be used to determine/predict the morphology of the RGs (which may suggest there are many physical processes that determine the radio morphology). 
It is also possible that the overlap in properties is due to
a mixture of RGs at different stages in their evolution, as we argued above.

We suspect the discrepancy between our finding and that of Owen et al.~-- who stated that the FR I and II RGs could be separated ``cleanly'' on the $P$--$M$ plane -- is due to sample construction. 
Substantial overlap between the two types is apparent
using either our sample or the subsample from \citet[][see Fig.~\ref{fig:config}]{gendre10}, both are radio flux-limited (see also \citealt{best09,wing10}).
However, sample selection was quite heterogeneous in some of the earlier works (OL89; \citealt{owen91,owen93}), where the main criterion for inclusion was to sample the $P$--$M$ plane as much as possible. In this sense, FR II objects are over-represented. 
With samples assembled under better defined criteria (e.g., limited by flux and redshift, restricted to central parts of clusters), some overlap between the two types were seen in \citet{ledlow96}.

%%%%%%%%%%%%%%%%%%%%%%%
\subsection{Origin of Different Morphologies?}
\label{sec:origin}

The existence of objects whose two lobes exhibit clearly different FR morphologies from each other
(``HYMORS'' objects, \citealt{gopal-krishna00}) argues that the immediate surrounding of the host galaxies must play some role in shaping the morphology.
In addition, analytic models of radio sources suggest that a
key quantity in determining the large scale morphology is the jet power-to-ambient density ratio ($L_j/\rho_a$) of the host galaxies (e.g., \citealt{kaiser07,kawakatu09}, and references therein). Here $\rho_a$ is measured at the core radius of the host galaxy.
Jets from hosts with low values of $L_j/\rho_a$ are more prone to the development of turbulence and become subsonic, resulting in plume-like morphology beyond the deceleration point, while jets from systems with high $L_j/\rho_a$ ratio are strong enough to remain supersonic, leading to the hot spots at the edge of the lobes.
Since the jet mechanical power is not directly observable, we assume it is proportional to $P_{1.4}^{0.7}$ \citep[e.g.,][and references therein]{cavagnolo10}. We also assume $\rho$ can be regarded as a faithful proxy for the local interstellar medium density in the host galaxies. The median mechanical power-to-total mass density ratio ($\propto P_{1.4}^{0.7}/\rho$) for $a_{\rm 0.9,em}$, $a_{\rm maj}$, and $b$ is roughly $3.2:2:1$, a trend in qualitative agreement with the models.
% median
In fact, the median values of $\rho$ for the three subsamples vary only by about 35\%, and it is mainly the difference in the radio power that drives the $P_{1.4}^{0.7}/\rho$ ratio.\footnote{The results are qualitatively the same if we use the $P_{1.4}^{0.7}/\mu$ ratio instead.}

% $5.1:2.3:1$
%
%(if using $\rho$ in place of $\mu$, the ratio is $3.4:2.3:1$)
%(if using $\mu$, the ratio is $4:2:1$)

We suspect that the accretion rate onto the SMBH is more important than the structure of the galaxy in determining the different radio morphologies 
%\citep[see also][]{baum95,hardcastle07,ho08}.
%%% NOTE!!!
(see also \citealt{baum95,ho08}; for discussions on the modes of accretion, see e.g., \citealt{best05b,hardcastle07,kauffmann08}).
At high accretion rates, the primary accretion flow is likely a geometrically thin, optically thick disk, which may launch jets that are very well-collimated over hundreds of kpc scale. As the accretion rate decreases, the thin disk moves away from the SMBH, and the inner region is occupied by a radiatively inefficient accretion flow (RIAF; see \citealt{esin97,narayan05}), with diminished emission line luminosity. Such an accretion flow is known to create outflows (e.g., \citealt{narayan95,blandford99}), and it is likely that jets so created will not be well-collimated (e.g., beyond hundreds of kpc),
%%% NOTE!!!
or are collimated initially, but suffer entrainment and deceleration very early on due to their lower intrinsic power.

If this picture is correct, we may understand the three subsets as follows. First, 
$a_{\rm 0.9,em}$ objects have highest SMBH accretion rates (e.g., OER $>10^{-6}$), are powered by classical thin accretion disks with strong, well-collimated jets that can produce strong hot spots, and are usually associated with lower mass galaxies living in less dense environments (with respect to the other subsamples considered here).
The $a_{\rm maj}$ RGs are massive, found in dense environments, and their central engines are likely fed by lower accretion rates (e.g., OER $<10^{-6}$), probably in a RIAF.
Finally, those galaxies with low accretion rates {\it and} with relatively low $L_j/\rho_a$ ratio will likely show jet-dominated morphology, making them class $b$ RGs.

To some degree the accretion rate correlates with $r_s$;
variations in the SMBH spin, magnetic fields, the structure of the galaxy, and the density of the intergalactic/intracluster medium, however, may all cause spreads in $r_s$ at a given accretion rate. It is possible that uncertainties in $M_{\rm BH}$ (inferred from the $M_{\rm BH}$--$\sigma$ relation) also smear the correlation.
About 12\% (9\%) of the $a_{<0.8}$ (class $b$) RGs have OER $\gtrsim 10^{-6}$;
while these high accretion rate class $b$ objects may represent earlier phases in evolution of CDs, the active class $a$ RGs with $r_s<0.8$ may be manifestations of the variations in $r_s$ at a given accretion rate mentioned above.

The difference in the accretion rate (and in turn the jet-launching mechanisms) may also explain the spectral properties of the host galaxies (see Figures~\ref{fig:derived} and \ref{fig:bhbins}).  For jets created by a thin disk, the ``zone of influence'' within the host galaxy is quite small (e.g., the jets may only punch two small ``holes'' in the galaxy), and thus any feedback due to the jets cannot suppress efficiently the star formation activity that may be linked to the onset of the AGN activity. On the other hand, if the jets launched by a RIAF are not well-collimated, they may influence a much larger volume of the host galaxy and thus terminate star formation more easily.  
The most efficient feedback mode (for the host galaxy itself) may be a combination of a RIAF and a dense interstellar medium (or immediate surrounding of the host galaxy), which slows down the jets quickly and creates the class $b$ morphology.

%%%%%%%%%%%%%%%%%%%%%%%%%%%%%%%%%%%%%%%%%%%%%%
\section{Conclusion}
\label{sec:conclusion}

Extended radio galaxies (RGs) have been classified based on their radio morphology or nuclear emission line activity. In this paper we have proposed a hybrid classification scheme that combines both features, and presented a comprehensive study of the host galaxy properties of RGs.
Our main objectives are to detect and define distinct populations of RGs, to understand the traditional Fanaroff-Riley (FR) type I/II dichotomy in the context of our new scheme, and to unravel the origin of different radio morphologies.
%
%We have presented a comprehensive study of the host galaxy properties of radio galaxies (RGs) of extended morphology, focusing on the Fanaroff-Riley (FR) type I/II dichotomy.
Our RG sample consists of \totalrg objects selected with 1.4 GHz radio flux density $f_{1.4}\ge3$ mJy, $r^{0.1}$-band absolute magnitude $M^{0.1}_r\le -21.27$ (i.e., more luminous than $M^*$, the characteristic magnitude of the galaxy luminosity function), radio angular diameter $T>30\arcsec$, physical size $T_p>40$ kpc, and at $0.02\le z\le 0.3$. All of the RGs in our sample appear to be massive early type galaxies.

We use the [\ion{O}{3}]$\lambda5007$ line luminosity as an indicator of the nuclear emission strength, and use a parameter $r_s\equiv S/T$ as a continuous parameterization of the RG radio morphology. Here $T$ is the total size of the radio sources, and $S$ is the separation between the highest surface brightness (HSB) spots on either sides of the galaxies.
%
%To better understand the relation between FR I and FR II, we measure the total size $T$ of the sources, and the separation $S$ between the highest surface brightness (HSB) spots on either sides of the galaxies; we then use the ratio $r_s\equiv S/T$ as a continuous parameterization of the RG populations. [Recall that in the binary FR classification scheme, type I (II) sources are defined with $r_s<0.5$ ($r_s>0.5$).]
%
Roughly 60\% of our objects show
HSB spots on both sides of the host galaxy; we refer to these as class $a$ RGs.
About 30\% of the sources appear to have prominent jets, with HSB spot coincident with the host galaxy. We call this population class $b$ (see Figures~\ref{fig:example} and \ref{fig:example2}; Table~\ref{tab:classes}).

Our main results are as follows:
\begin{itemize}
\item The distribution of $r_s$ is bimodal (Fig.~\ref{fig:sthist}), although we argue that the two peaks do not correspond to the two FR types (\S\S\ref{sec:morph},\ref{sec:phys_prop}).

\item Among the class $a$ objects, a small population with high values of $r_s$ ($\gtrsim 0.8$) {\it and} high [\ion{O}{3}]$\lambda5007$ line luminosity ($L_{\rm OIII}>10^6 L_\odot$) seems to be distinguished from the rest, in the sense that {\it on average} they are hosted by lower mass galaxies, live in relatively sparse environments, and have higher accretion rates onto the central supermassive black hole (SMBH), as manifested by the [\ion{O}{3}]$\lambda5007$ line Eddington ratio (\S\ref{sec:classa}; Figures~\ref{fig:gal}--\ref{fig:rad}). We refer to these RGs as the $a_{\rm 0.9,em}$ subset, and the rest, the majority of class $a$, as the $a_{\rm maj}$ subset (Table~\ref{tab:subclasses}). The distribution of $r_s$ for the $a_{\rm maj}$ objects is quite broad, encompassing the range occupied by the $a_{\rm 0.9,em}$ RGs at the high $r_s$ end.
A simple morphological measure such as $r_s$ is thus only of limited use for classifying extended RGs.

\item The average properties of  $a_{\rm maj}$ and class $b$ RGs, such as the optical luminosity, stellar mass, 4000\AA\ break strength, and velocity dispersion, differ by 20\% or less from one another. However, because of the differences in the environments (e.g., characterized by the neighbor counts within 0.5 Mpc; Fig.~\ref{fig:rad}, panel b) and the (nuclear) emission line properties (Fig.~\ref{fig:rad}, panels g \& h; Table~\ref{tab:agn}), we regard them as distinct populations (\S\S\ref{sec:classb},\ref{sec:rlvsrq}). 
\item Among the three subsamples ($a_{\rm 0.9,em}$, $a_{\rm maj}$, $b$), galaxies in class $b$ have the lowest Eddington ratio and radio power, and their nuclear and/or star formation activity even appears to be suppressed relative to the radio quiet (RQ) galaxies that have similar luminosities and mass (\S\ref{sec:rlvsrq}; Table~\ref{tab:agn}).

\item Different researchers usually have adopted somewhat different definitions for the FR I/II types. As our proposed classification scheme is based on both radio morphology and nuclear emission line strength, and the original FR scheme is purely morphology-based, 
there is no one-to-one correspondence between the two.
%our RG subsets and the type I/II based on the original FR definition, which is purely morphology-based.
Nevertheless, given the similarities of class $b$ and class $a$ objects with $r_s\le 0.8$, and the large difference between these subsamples and the class $a$ RGs with $r_s>0.8$ (\S\ref{sec:classa}), we can broadly identify the FR I type with the former and FR II with the latter sources (see the discussion in \S\ref{sec:threetypes}).
However, there is considerable overlap in the distributions of physical properties for the three subsamples, and the transition from one FR type to the other is far from sharp (\S\S\ref{sec:owen},\ref{sec:mpagain}).
In particular, our findings do not support the previous claim that the two FR types occupy distinct regions in the radio luminosity-optical magnitude plane.

\item Although on average the $a_{\rm 0.9,em}$ objects are less massive than the other RG subsamples, they are still hosted by massive galaxies ($M_r^{0.1}\lesssim M^*$).  The distinction of this subset from the other RGs is far less dramatic than the blue--red bimodality of the general galaxy population.

\item To single out $a_{\rm 0.9,em}$ from the rest of the population in a statistically complete, low redshift RG sample for which optical emission line measurements are not available, a possible approach is to select sources with $r_s$ 
ranked in the top 10\% of the distribution.

\item Some of the objects in class $b$ may form an evolutionary sequence, that is, they can be regarded as RGs seen at different stages of evolution, as evidenced by an anti-correlation between size and $r_s$ (\S\ref{sec:evol}; Fig.~\ref{fig:ev}). 
A larger sample is needed to evaluate the effect of systematic uncertainties in the sample selection, however.

\item Many different mechanisms must be at work for the generation of radio jets, but we suggest that the accretion rate onto the SMBH is the main driver for the different radio morphologies, with host galaxy structure and/or density of the surrounding environment playing a secondary role.
This is primarily motivated by the stark differences in the nuclear emission properties of the RG subsets.
At high accretion rates (e.g., [\ion{O}{3}]$\lambda5007$ line Eddington ratio $>10^{-6}$), the accretion mode is likely dominated by a geometrically thin, optically thick disk which
could generate powerful, well-collimated jets that create strong hot spots at the edge of the lobes (i.e., the ``classical double'' morphology). At lower accretion rates, a radiatively inefficient accretion flow (RIAF) takes over; the outflows/jets from such accretion flows may not be as well-collimated 
%%% NOTE!!!
beyond the galactic nucleus scale,
%%% NOTE!!!
resulting in the ``fat double'' morphology (i.e., without obvious hot spots at the edge of the lobes). At slightly lower accretion rates (e.g., [\ion{O}{3}] Eddington ratio $\lesssim 10^{-7}$) {\it and} for galaxies with sufficiently high galactic density, a jet-dominated morphology is created (\S\ref{sec:origin}). 

\item Based on the spectral properties of the galaxies, we suggest that outflows/jets from a RIAF may be more efficient in suppressing processes that cause star formation and/or nuclear activity than the jets from thin accretion disks. The latter could affect the large scale surroundings of the RGs, however (\S\ref{sec:origin}, Fig.~\ref{fig:rad}).

\end{itemize}

The advent of wide-field, uniform radio and optical surveys such as NVSS, FIRST, and SDSS makes it possible to produce the large RG sample used here, and the classification scheme we propose. Although it is not clear if our scheme is more physically motivated than the existing ones (e.g., those of FR and OL89), our classification should be among the most objective and quantitative, and easily reproducible by other researchers.

In this study we have only concerned ourselves with the extended sources with relatively ``straight'' lobes, that is, we have excluded wide-angle tail and narrow-angle tail objects. We have also left out the compact/point-like sources and radio quasars in the analysis. In a future publication we will compare the host properties of RGs of these other morphologies, which may provide further insights into the generation of the radio emission in galactic nuclei.

%%%%%%%%%%%%%%%%%%%%%%%%%%%%%%%%%%%%%%%%%%%%%%
\acknowledgments

We thank the referee, Philip Best, for an insightful, careful, and encouraging report that improved the clarity and presentation of the paper.
We are grateful to John Silverman, Jim Gunn, Nozomu Kawakatu, Jonghak Woo, Ron Taam, Paul Wiita, Rick White, Jim Condon, Brian Mason, Sheng-Yuan Liu, Jarle Brinchmann, Wei-Hao Wang, and Melanie Gendre for helpful discussions and comments, and to Robert Lupton for help with sm.
YTL thanks IH for constant support and inspiration.
YTL acknowledges supports from 
the World Premier International Research Center Initiative, MEXT, Japan,
and 
from a Princeton-Cat\'{o}lica Fellowship and NSF PIRE grant OISE-0530095, while he was at Princeton.
YS acknowledges support from a Clay Postdoctoral 
Fellowship through the Smithsonian Astrophysical Observatory.
MAS and YS acknowledge the support of NSF grant AST-0707266.
GTR was supported in part by an Alfred P.~Sloan Research Fellowship.

Funding for the Sloan Digital Sky Survey (SDSS) and SDSS-II has been provided
by the Alfred P. Sloan Foundation, the Participating Institutions, the National
Science Foundation, the U.S. Department of Energy, the National Aeronautics and
Space Administration, the Japanese Monbukagakusho, and the Max Planck Society,
and the Higher Education Funding Council for England. The SDSS Web site is
http://www.sdss.org/.

This research has made use of the NED database, the data products
from the NVSS and FIRST surveys, and the Aladin sky atlas. The extensive support from the CDS helpdesk is much appreciated.

%\vspace{-4mm}

%%%%%%%%%%%%%%%%%%%%%%%%%%%%%%%%%%%%%%%%%%%%%%
%%%%%%%%%%%%%%%%%%%%%%%%%%%%%%%%%%%%%%%%%%%%%%
%%%%%%%%%%%%%%%%%%%%%%%%%%%%%%%%%%%%%%%%%%%%%%
\appendix
\label{sec:app}

%\vspace{-6mm}

We noted in \S\ref{sec:morph} that the measurements for both $S$ and $T$ may depend on the sensitivity and resolution of the radio data. We show below (Appendix \ref{sec:first}) that the FIRST-based measurement of $S$ should be robust for our RG sample.
Our main results are also shown to be qualitatively insensitive to the data (FIRST or NVSS) from which total size $T$ is measured (Appendix \ref{sec:totsize}).

\vspace{+2mm}

%%%%%%%%%%%%%%%%%%%%%%%%%%%%%%%%%%%%%%%%%%%%%%
\section{Resolution of FIRST Images}
\label{sec:first}

Part of the SDSS Stripe 82 has been observed with the VLA in the A configuration ($1.5\arcsec$ FWHM) at 1.4 GHz, reaching rms $\sim 0.07$ mJy/beam.  15 RGs in our sample lie in the region covered by this deep survey\footnote{\url{http://www.physics.drexel.edu/$\sim$gtr/vla/stripe82/}} (PI: G.~Richards; Hodge et al.~2010, in preparation). Their redshifts range from 0.04 to 0.25, which is representative of our sample (Fig.~\ref{fig:gal}, panel j). We compare the $S$ and $T$ measurements from these data with those from FIRST (hereafter with the subscripts $V$ and $F$, respectively).  Among the 15 objects, five are in class $b$. For the ten class $a$ RGs, the ratio $S_V/S_F$ has a mean of 0.95 and a scatter of 0.15. We also find that the total size measurements are very close, which suggests that the $r_s$ value derived from FIRST is robust against resolution issues.

For four of the class $a$ RGs, the HSB coincides with prominent jets in the host galaxy in the deep VLA images, and are classified as class $b$ or $c$, depending on the central-to-total flux ratio (\S\ref{sec:morph}). We note, however, that some of the flux from the lobes is resolved out in these A-array maps, and therefore ideally one should combine both the high and low resolution data to measure the proportion of flux that is in the jet component which only shows up in high resolution maps.
These objects tend to have $r_f$ values higher than the majority of the class $a$ RGs, based on FIRST data; that is, FIRST does detect the central component. Because of the poorer angular resolution of FIRST, the jets are not as prominent as in the A-array maps.

For the current analysis, we acknowledge the possibility that some of our objects which we have put into class $a$ may in fact belong to class $b$ if measured with better data. Using the distribution of (FIRST-based) $r_s$ and $r_f$ of the above four class $a$ RGs, we estimate that 13\% of class $a$ objects may be subject to this misclassification.

Although the redshift distribution for class $a$ is quite consistent across the $r_s$ bins, class $b$ RGs with higher $r_s$ are on average at slightly higher redshift (Fig.~\ref{fig:gal}, panel j). Given the inherent sensitivity of the classification on the resolution, this is perhaps not surprising. For the five class $b$ objects with deeper, higher-resolution VLA data, two have $r_s \approx 0.5$ based on FIRST, and have redshift of 0.224 and 0.252, respectively. At three times finer resolution than FIRST, these RGs remain jet-dominated (i.e., class $b$), although their $r_s$ decreases. Given that the properties of class $b$ objects do not vary much with respect to $r_s$, we conclude that our results for class $b$ should be robust (except for the possible addition of class $a$ RGs with better measurements).

\begin{figure*}
\epsscale{0.77}
\plotone{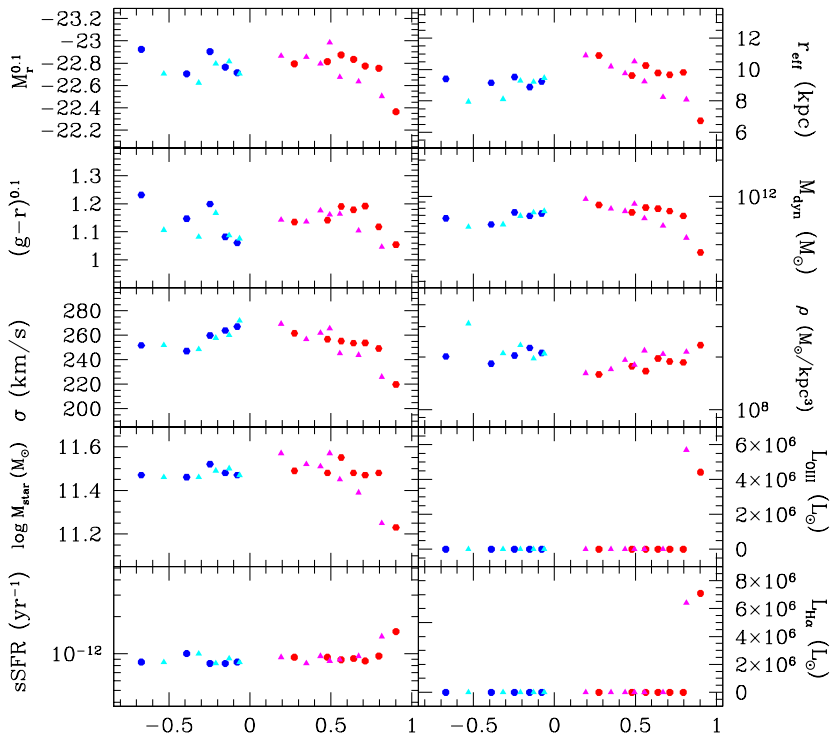}
\vspace{-10mm}
\caption{
Median values of several physical properties as a function of $r_s$ (c.f.~Fig.~\ref{fig:gal} and Fig.~\ref{fig:derived}),
with $T$ derived from NVSS (magenta and cyan triangles) and those with $T$ from FIRST (red and blue points). For class $b$ RGs (blue and cyan points), we plot the results with negative values of $r_s$ to avoid cluttering the Figure.
}
\label{fig:first4}
\end{figure*}

\begin{figure*}
\epsscale{0.77}
\plotone{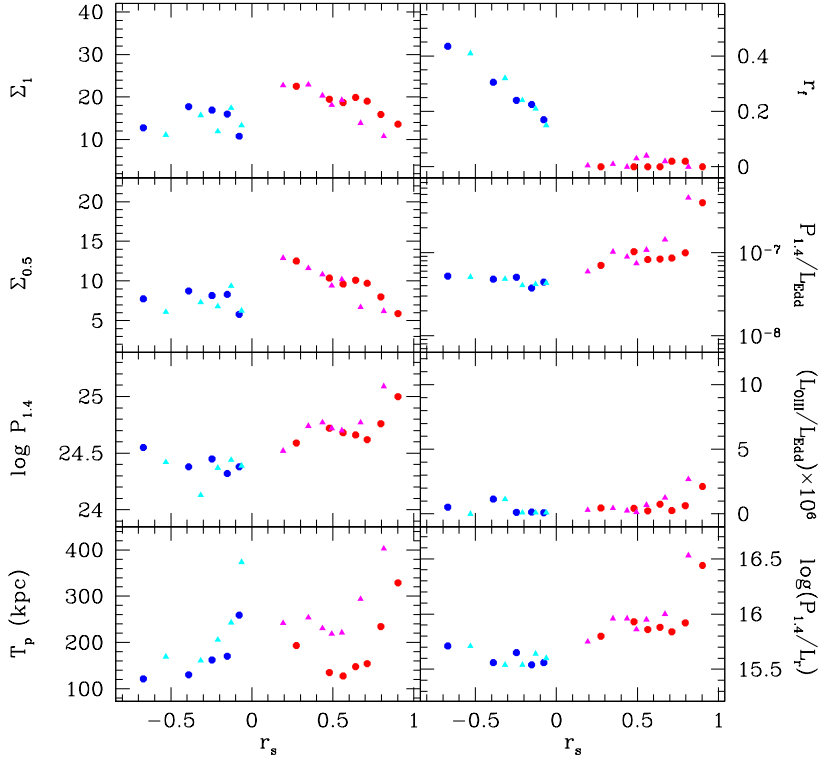}
\vspace{-10mm}
\caption{
Similar to Fig.~\ref{fig:rad}, but showing median results with $T$ derived from NVSS (magenta and cyan triangles) and those with $T$ from FIRST (red and blue points). For class $b$ RGs (blue and cyan points), we plot the results with negative values of $r_s$ to avoid cluttering the Figure.
}
\label{fig:first6}
\end{figure*}

%%%%%%%%%%%%%%%%%%%%%%%%%%%%%%%%%%%%%%%%%%%%%%
\section{Measurement of Total Size}
\label{sec:totsize}

We have the option of using either NVSS or FIRST data to measure the total size $T$ of the radio sources, and have chosen to use the latter as the default (except for the 41 cases where the FIRST-based sizes are much less than those from NVSS, presumably due to the insensitivity of FIRST to diffuse emission; see \S\ref{sec:morph}). 
We have repeated our analysis with NVSS-based $T$ measurements.
If the fitted size of the major axis of radio sources from NVSS is only an upper limit, we exclude the sources from the sample; in addition, a minimal size of $T=50\arcsec$ (rather than $T=30\arcsec$ as adopted in \S\ref{sec:morph}) is imposed, and therefore the sample size (797 RGs) with NVSS-based measurement is smaller.

Variations of physical properties as a function of $r_s$, analogous to Figures~\ref{fig:gal}--\ref{fig:rad}, are shown in Figures~\ref{fig:first4} and \ref{fig:first6}, for both FIRST-based and NVSS-based results.
The main difference is in the total size $T_p$ of the sources (Fig.~\ref{fig:first6}, lower left panel): those derived from NVSS are larger than the ones based on FIRST, as expected. 
Since $S$ is still measured using FIRST data and thus remains unchanged,
the NVSS-based $r_s$ values are systematically lower than the FIRST-based ones; 
any trends with $r_s$ seen in Figures~\ref{fig:gal}--\ref{fig:rad} would therefore appear ``stretched'' horizontally and shifted towards low $r_s$ a bit.

Using NVSS-based $T$ measurements, we still find that class $a$ objects with the highest $r_s$ stand out from the rest of the sample, although the division is now at $r_{s}\approx 0.7$.
The subtle difference in properties between class $b$ and class $a$ RGs with $r_s<0.7$ also remains.

A possible concern of using either NVSS or FIRST to measure $T$ is the relatively high SB limits of these surveys. One could imagine that a FD source with HSB spots far from the edge of the lobes would appear as high $r_s$ objects if observed with insufficient depth.
We have checked against the NASA/IPAC Extragalactic Database (NED) to look for archival radio images for our $a_{0.9}$ objects. Only 8 RGs ($\sim 10\%$) have been imaged with decent data from the literature, and all of them would still have high $r_s$ in those deeper maps.
Even if such a bias due to the depth of the surveys exists, we suspect the difference in morphology between a bona fide hot spot at the edge of a lobe and a HSB region within a lobe for lower-$r_s$ RGs would be obvious enough in FIRST images so that a visual inspection would be able to pick up such cases.

We noted in \S\ref{sec:threetypes} that the combination of the presence of hot spots at the edge of the lobes {\it and} the high accretion rate as indicated by the presence of emission lines seems to be a pretty robust indicator for the type of RGs corresponding to FR II, or the CDs.
Since $r_s$ is unfortunately somewhat resolution-dependent, and emission line properties in the optical are not always easily available,
perhaps a more objective approach to single out CDs from the rest of the population in a statistically complete, low redshift RG sample (so that our results are fully applicable)
is to select sources with $r_s$ 
ranked in the top 10\% 
of the distribution.

%%% NOTE: need to delete the last \\ in the data body

\begin{deluxetable}{lrcrrrrrrrrrrr}
%\tighten 

\tabletypesize{\scriptsize}
\tablecaption{Basic properties of Radio Galaxies} 
%\tablewidth{0pt}

\tablehead{
  \colhead{SDSS ID} & 
  \colhead{VAGC ID\tablenotemark{a}} & 
  \colhead{Class} & 
  \colhead{RA} & 
  \colhead{Dec} & 
  \colhead{$z$} & 
  \colhead{$\log\,P_{1.4}$} & 
  \colhead{$M_r^{0.1}$} & 
  \colhead{$\sigma$} &
  \colhead{$r_{\rm eff}$} &
  \colhead{$T_p$} &
  \colhead{$r_s$} &
  \colhead{$L_{\rm OIII}$} &
  \colhead{$\Sigma_{0.5}$} \\
  \colhead{} &
  \colhead{} &
  \colhead{} &
  \colhead{(J2000)} &
  \colhead{(J2000)} &
  \colhead{} &
  \colhead{(W/Hz)} &
  \colhead{} &
  \colhead{(km/s)} &
  \colhead{(kpc)} &
  \colhead{(kpc)} &
  \colhead{} &
  \colhead{($10^6 L_\odot$)} &
  \colhead{} 
}

\startdata

587731186743967908 & 1659866 & $b$ &   7.247278 & $  0.333494$ &  0.2222 & 24.44 & $-23.063$ & 331 & 11.5 &  111 & 0.48 & \nodata & $22.9$\\
588015510345613463 & 1699168 & $b$ &   5.794018 & $  0.976323$ &  0.2270 & 24.68 & $-23.039$ & 266 & 9.2 &  120 & 0.24 & \nodata & $15.6$\\
587724232636498080 & 354077 & $a$ &   7.205604 & $ 14.979081$ &  0.0977 & 24.96 & $-21.627$ & 177 & 6.8 &  368 & 0.56 & \nodata & $ 1.4$\\
587731185132503151 & 1652016 & $a$ &   5.281742 & $ -0.925413$ &  0.1082 & 24.52 & $-22.985$ & 268 & 9.5 &  117 & 0.69 & \nodata & $14.2$

\enddata

\tablecomments{The full table is available and kept up-to-date at http://member.ipmu.jp/yen-ting.lin/RG/index.html}

\tablenotetext{a}{NYU VAGC DR6 object ID.}

\label{tab:maintable}

\end{deluxetable}

%\input{../newmaintable}

%%%%%%%%%%%%%%%%%%%%%%%%%%%%%%%%%%%%%%%%%%%%%%
%\bibliographystyle{apj}
%\bibliography{cosmology,refs}

\begin{thebibliography}{59}
\expandafter\ifx\csname natexlab\endcsname\relax\def\natexlab#1{#1}\fi

\bibitem[{{Adelman-McCarthy} {et~al.}(2008){Adelman-McCarthy}, {Ag{\"u}eros},
  {Allam}, {Allende Prieto}, {Anderson}, {Anderson}, {Annis}, {Bahcall},
  {Bailer-Jones}, {Baldry}, {Barentine}, {Bassett}, {Becker}, {Beers}, {Bell},
  {Berlind}, {Bernardi}, {Blanton}, {Bochanski}, {Boroski}, {Brinchmann},
  {Brinkmann}, {Brunner}, {Budav{\'a}ri}, {Carliles}, {Carr}, {Castander},
  {Cinabro}, {Cool}, {Covey}, {Csabai}, {Cunha}, {Davenport}, {Dilday}, {Doi},
  {Eisenstein}, {Evans}, {Fan}, {Finkbeiner}, {Friedman}, {Frieman},
  {Fukugita}, {G{\"a}nsicke}, {Gates}, {Gillespie}, {Glazebrook}, {Gray},
  {Grebel}, {Gunn}, {Gurbani}, {Hall}, {Harding}, {Harvanek}, {Hawley},
  {Hayes}, {Heckman}, {Hendry}, {Hindsley}, {Hirata}, {Hogan}, {Hogg}, {Hyde},
  {Ichikawa}, {Ivezi{\'c}}, {Jester}, {Johnson}, {Jorgensen}, {Juri{\'c}},
  {Kent}, {Kessler}, {Kleinman}, {Knapp}, {Kron}, {Krzesinski}, {Kuropatkin},
  {Lamb}, {Lampeitl}, {Lebedeva}, {Lee}, {Leger}, {L{\'e}pine}, {Lima}, {Lin},
  {Long}, {Loomis}, {Loveday}, {Lupton}, {Malanushenko}, {Malanushenko},
  {Mandelbaum}, {Margon}, {Marriner}, {Mart{\'{\i}}nez-Delgado}, {Matsubara},
  {McGehee}, {McKay}, {Meiksin}, {Morrison}, {Munn}, {Nakajima}, {Neilsen},
  {Newberg}, {Nichol}, {Nicinski}, {Nieto-Santisteban}, {Nitta}, {Okamura},
  {Owen}, {Oyaizu}, {Padmanabhan}, {Pan}, {Park}, {Peoples}, {Pier}, {Pope},
  {Purger}, {Raddick}, {Re Fiorentin}, {Richards}, {Richmond}, {Riess}, {Rix},
  {Rockosi}, {Sako}, {Schlegel}, {Schneider}, {Schreiber}, {Schwope}, {Seljak},
  {Sesar}, {Sheldon}, {Shimasaku}, {Sivarani}, {Smith}, {Snedden}, {Steinmetz},
  {Strauss}, {SubbaRao}, {Suto}, {Szalay}, {Szapudi}, {Szkody}, {Tegmark},
  {Thakar}, {Tremonti}, {Tucker}, {Uomoto}, {Vanden Berk}, {Vandenberg},
  {Vidrih}, {Vogeley}, {Voges}, {Vogt}, {Wadadekar}, {Weinberg}, {West},
  {White}, {Wilhite}, {Yanny}, {Yocum}, {York}, {Zehavi}, \&
  {Zucker}}]{sdssdr6}
{Adelman-McCarthy}, J.~K., et~al.
  2008, \apjs, 175, 297
  
\bibitem[Baldry et al.(2004)]{baldry04} Baldry, I.~K., 
Glazebrook, K., Brinkmann, J., Ivezi{\'c}, {\v Z}., Lupton, R.~H., Nichol, 
R.~C., \& Szalay, A.~S.\ 2004, \apj, 600, 681 



%\bibitem[{{B{\^ i}rzan} {et~al.}(2004){B{\^ i}rzan}, {Rafferty}, {McNamara},
%  {Wise}, \& {Nulsen}}]{birzan04}
%{B{\^ i}rzan}, L., {Rafferty}, D.~A., {McNamara}, B.~R., {Wise}, M.~W., \&
%  {Nulsen}, P.~E.~J. 2004, \apj, 607, 800

\bibitem[{{Baldi} \& {Capetti}(2010)}]{baldi10}
{Baldi}, R.~D. \& {Capetti}, A.\ 2010, \aap, accepted (arXiv:1005.3223)


\bibitem[{{Baldwin} {et~al.}(1981){Baldwin}, {Phillips}, \&
  {Terlevich}}]{baldwin81}
{Baldwin}, J.~A., {Phillips}, M.~M., \& {Terlevich}, R. 1981, \pasp, 93, 5

\bibitem[Barthel(1989)]{barthel89} Barthel, P.~D.\ 1989, \apj, 
336, 606 



\bibitem[Baum \& Heckman(1989)]{baum89} Baum, S.~A., \& Heckman, T.\ 1989, \apj, 336, 681 



\bibitem[{{Baum} {et~al.}(1995){Baum}, {Zirbel}, \& {O'Dea}}]{baum95}
{Baum}, S.~A., {Zirbel}, E.~L., \& {O'Dea}, C.~P. 1995, \apj, 451, 88

\bibitem[{{Becker} {et~al.}(1995){Becker}, {White}, \& {Helfand}}]{becker95}
{Becker}, R.~H., {White}, R.~L., \& {Helfand}, D.~J. 1995, \apj, 450, 559

\bibitem[{{Best} {et~al.}(2005a){Best}, {Kauffmann}, {Heckman}, \&
  {Ivezi{\'c}}}]{best05a}
{Best}, P.~N., {Kauffmann}, G., {Heckman}, T.~M., \& {Ivezi{\'c}}, {\v Z}.
  2005a, \mnras, 362, 9

\bibitem[Best et al.(2005b)]{best05b} Best, P.~N., Kauffmann, 
G., Heckman, T.~M., Brinchmann, J., Charlot, S., Ivezi{\'c}, {\v Z}., 
\& White, S.~D.~M.\ 2005b, \mnras, 362, 25 



  
\bibitem[Best(2009)]{best09} Best, P.~N.\ 2009, Astronomische 
Nachrichten, 330, 184 



\bibitem[{{Blandford} \& {Begelman}(1999)}]{blandford99}
{Blandford}, R.~D. \& {Begelman}, M.~C. 1999, \mnras, 303, L1

\bibitem[{{Blanton} {et~al.}(2003){Blanton}, {Hogg}, {Bahcall}, {Brinkmann},
  {Britton}, {Connolly}, {Csabai}, {Fukugita}, {Loveday}, {Meiksin}, {Munn},
  {Nichol}, {Okamura}, {Quinn}, {Schneider}, {Shimasaku}, {Strauss}, {Tegmark},
  {Vogeley}, \& {Weinberg}}]{blanton03b}
{Blanton}, M.~R., {Hogg}, D.~W., {Bahcall}, N.~A., {Brinkmann}, J., {Britton},
  M., {Connolly}, A.~J., {Csabai}, I., {Fukugita}, M., {Loveday}, J.,
  {Meiksin}, A., {Munn}, J.~A., {Nichol}, R.~C., {Okamura}, S., {Quinn}, T.,
  {Schneider}, D.~P., {Shimasaku}, K., {Strauss}, M.~A., {Tegmark}, M.,
  {Vogeley}, M.~S., \& {Weinberg}, D.~H. 2003, \apj, 592, 819

\bibitem[{{Blanton} {et~al.}(2005){Blanton}, {Schlegel}, {Strauss},
  {Brinkmann}, {Finkbeiner}, {Fukugita}, {Gunn}, {Hogg}, {Ivezi{\'c}}, {Knapp},
  {Lupton}, {Munn}, {Schneider}, {Tegmark}, \& {Zehavi}}]{blanton05}
{Blanton}, M.~R., {Schlegel}, D.~J., {Strauss}, M.~A., {Brinkmann}, J.,
  {Finkbeiner}, D., {Fukugita}, M., {Gunn}, J.~E., {Hogg}, D.~W., {Ivezi{\'c}},
  {\v Z}., {Knapp}, G.~R., {Lupton}, R.~H., {Munn}, J.~A., {Schneider}, D.~P.,
  {Tegmark}, M., \& {Zehavi}, I. 2005, \aj, 129, 2562

\bibitem[{{Bonnarel} {et~al.}(2000){Bonnarel}, {Fernique}, {Bienaym{\'e}},
  {Egret}, {Genova}, {Louys}, {Ochsenbein}, {Wenger}, \&
  {Bartlett}}]{bonnarel00}
{Bonnarel}, F., {Fernique}, P., {Bienaym{\'e}}, O., {Egret}, D., {Genova}, F.,
  {Louys}, M., {Ochsenbein}, F., {Wenger}, M., \& {Bartlett}, J.~G. 2000,
  \aaps, 143, 33

\bibitem[{{Brinchmann} {et~al.}(2004){Brinchmann}, {Charlot}, {White},
  {Tremonti}, {Kauffmann}, {Heckman}, \& {Brinkmann}}]{brinchmann04}
{Brinchmann}, J., {Charlot}, S., {White}, S.~D.~M., {Tremonti}, C.,
  {Kauffmann}, G., {Heckman}, T., \& {Brinkmann}, J. 2004, \mnras, 351, 1151


\bibitem[Cavagnolo et al.(2010)]{cavagnolo10} Cavagnolo, K.~W., 
McNamara, B.~R., Nulsen, P.~E.~J., Carilli, C.~L., Jones, C., 
\& Birzan, L.\ 2010, \apj, in press (arXiv:1006.5699) 



\bibitem[{{Condon} {et~al.}(1998){Condon}, {Cotton}, {Greisen}, {Yin},
  {Perley}, {Taylor}, \& {Broderick}}]{condon98}
{Condon}, J.~J., {Cotton}, W.~D., {Greisen}, E.~W., {Yin}, Q.~F., {Perley},
  R.~A., {Taylor}, G.~B., \& {Broderick}, J.~J. 1998, \aj, 115, 1693

\bibitem[De Young(1993)]{deyoung93} De Young, D.~S.\ 1993, \apjl, 
405, L13 

\bibitem[{{De Young}(2002)}]{deyoung02}
{De Young}, D.~S. 2002, {The physics of extragalactic radio sources}
  (University of Chicago Press)

\bibitem[{{Esin} {et~al.}(1997){Esin}, {McClintock}, \& {Narayan}}]{esin97}
{Esin}, A.~A., {McClintock}, J.~E., \& {Narayan}, R. 1997, \apj, 489, 865

\bibitem[{{Falcke} \& {Biermann}(1995)}]{falcke95}
{Falcke}, H. \& {Biermann}, P.~L. 1995, \aap, 293, 665

\bibitem[{{Fanaroff} \& {Riley}(1974)}]{fanaroff74}
{Fanaroff}, B.~L. \& {Riley}, J.~M. 1974, \mnras, 167, 31P

\bibitem[{{Gehrels}(1986)}]{gehrels86}
{Gehrels}, N. 1986, \apj, 303, 336

\bibitem[{{Gendre} {et~al.}(2010){Gendre}, {Best}, \& {Wall}}]{gendre10}
{Gendre}, M.~A., {Best}, P.~N., \& {Wall}, J.~V. 2010, \mnras, 404, 1719

\bibitem[{{Gendre} \& {Wall}(2008)}]{gendre08}
{Gendre}, M.~A. \& {Wall}, J.~V. 2008, \mnras, 390, 819

\bibitem[{{Gopal-Krishna} \& {Wiita}(2000)}]{gopal-krishna00}
{Gopal-Krishna} \& {Wiita}, P.~J. 2000, \aap, 363, 507

\bibitem[{{Graham} {et~al.}(2005){Graham}, {Driver}, {Petrosian}, {Conselice},
  {Bershady}, {Crawford}, \& {Goto}}]{graham05}
{Graham}, A.~W., {Driver}, S.~P., {Petrosian}, V., {Conselice}, C.~J.,
  {Bershady}, M.~A., {Crawford}, S.~M., \& {Goto}, T. 2005, \aj, 130, 1535
  
\bibitem[Hardcastle et al.(2006)]{hardcastle06} Hardcastle, M.~J., 
Evans, D.~A., \& Croston, J.~H.\ 2006, \mnras, 370, 1893 
  

\bibitem[{{Hardcastle} {et~al.}(2007){Hardcastle}, {Evans}, \&
  {Croston}}]{hardcastle07}
{Hardcastle}, M.~J., {Evans}, D.~A., \& {Croston}, J.~H. 2007, \mnras, 376,
  1849

\bibitem[{{Heckman} {et~al.}(1994){Heckman}, {O'Dea}, {Baum}, \&
  {Laurikainen}}]{heckman94}
{Heckman}, T.~M., {O'Dea}, C.~P., {Baum}, S.~A., \& {Laurikainen}, E. 1994,
  \apj, 428, 65


  
%\bibitem[Herbert et al.(2010)]{herbert10} Herbert, P.~D., Jarvis, 
%M.~J., Willott, C.~J., McLure, R.~J., Mitchell, E., Rawlings, S., Hill, 
%G.~J., \& Dunlop, J.~S.\ 2010, \mnras, 406, 1841 


  
\bibitem[Heywood et al.(2007)]{heywood07} Heywood, I., Blundell, 
K.~M., \& Rawlings, S.\ 2007, \mnras, 381, 1093 

  

\bibitem[{{Hine} \& {Longair}(1979)}]{hine79}
{Hine}, R.~G. \& {Longair}, M.~S. 1979, \mnras, 188, 111

\bibitem[{{Ho}(2008)}]{ho08}
{Ho}, L.~C. 2008, \araa, 46, 475

\bibitem[{{Ivezi{\'c}} {et~al.}(2002){Ivezi{\'c}}, {Menou}, {Knapp}, {Strauss},
  {Lupton}, {Vanden Berk}, {Richards}, {Tremonti}, {Weinstein}, {Anderson},
  {Bahcall}, {Becker}, {Bernardi}, {Blanton}, {Eisenstein}, {Fan},
  {Finkbeiner}, {Finlator}, {Frieman}, {Gunn}, {Hall}, {Kim}, {Kinkhabwala},
  {Narayanan}, {Rockosi}, {Schlegel}, {Schneider}, {Strateva}, {SubbaRao},
  {Thakar}, {Voges}, {White}, {Yanny}, {Brinkmann}, {Doi}, {Fukugita},
  {Hennessy}, {Munn}, {Nichol}, \& {York}}]{ivezic02}
{Ivezi{\'c}}, {\v Z}., et al.
  2002, \aj, 124, 2364

\bibitem[{{Kaiser} \& {Best}(2007)}]{kaiser07}
{Kaiser}, C.~R. \& {Best}, P.~N. 2007, \mnras, 381, 1548

\bibitem[{{Kauffmann} {et~al.}(2008){Kauffmann}, {Heckman}, \&
  {Best}}]{kauffmann08}
{Kauffmann}, G., {Heckman}, T.~M., \& {Best}, P.~N. 2008, \mnras, 384, 953

\bibitem[{{Kauffmann} {et~al.}(2003{\natexlab{a}}){Kauffmann}, {Heckman},
  {Tremonti}, {Brinchmann}, {Charlot}, {White}, {Ridgway}, {Brinkmann},
  {Fukugita}, {Hall}, {Ivezi{\' c}}, {Richards}, \& {Schneider}}]{kauffmann03}
{Kauffmann}, G., {Heckman}, T.~M., {Tremonti}, C., {Brinchmann}, J., {Charlot},
  S., {White}, S.~D.~M., {Ridgway}, S.~E., {Brinkmann}, J., {Fukugita}, M.,
  {Hall}, P.~B., {Ivezi{\' c}}, {\v Z}., {Richards}, G.~T., \& {Schneider},
  D.~P. 2003{\natexlab{a}}, \mnras, 346, 1055

\bibitem[{{Kauffmann} {et~al.}(2003{\natexlab{b}}){Kauffmann}, {Heckman},
  {White}, {Charlot}, {Tremonti}, {Peng}, {Seibert}, {Brinkmann}, {Nichol},
  {SubbaRao}, \& {York}}]{kauffmann03b}
{Kauffmann}, G., {Heckman}, T.~M., {White}, S.~D.~M., {Charlot}, S.,
  {Tremonti}, C., {Peng}, E.~W., {Seibert}, M., {Brinkmann}, J., {Nichol},
  R.~C., {SubbaRao}, M., \& {York}, D. 2003{\natexlab{b}}, \mnras, 341, 54

\bibitem[{{Kawakatu} {et~al.}(2009){Kawakatu}, {Kino}, \& {Nagai}}]{kawakatu09}
{Kawakatu}, N., {Kino}, M., \& {Nagai}, H. 2009, \apjl, 697, L173

\bibitem[{{Kimball} \& {Ivezi{\'c}}(2008)}]{kimball08}
{Kimball}, A.~E. \& {Ivezi{\'c}}, {\v Z}. 2008, \aj, 136, 684

\bibitem[{{Koester} {et~al.}(2007){Koester}, {McKay}, {Annis}, {Wechsler},
  {Evrard}, {Bleem}, {Becker}, {Johnston}, {Sheldon}, {Nichol}, {Miller},
  {Scranton}, {Bahcall}, {Barentine}, {Brewington}, {Brinkmann}, {Harvanek},
  {Kleinman}, {Krzesinski}, {Long}, {Nitta}, {Schneider}, {Sneddin}, {Voges},
  \& {York}}]{koester07}
{Koester}, B.~P., {McKay}, T.~A., {Annis}, J., {Wechsler}, R.~H., {Evrard}, A.,
  {Bleem}, L., {Becker}, M., {Johnston}, D., {Sheldon}, E., {Nichol}, R.,
  {Miller}, C., {Scranton}, R., {Bahcall}, N., {Barentine}, J., {Brewington},
  H., {Brinkmann}, J., {Harvanek}, M., {Kleinman}, S., {Krzesinski}, J.,
  {Long}, D., {Nitta}, A., {Schneider}, D.~P., {Sneddin}, S., {Voges}, W., \&
  {York}, D. 2007, \apj, 660, 239
%{Koester}, B.~P., et al.~2007, \apj, 660, 239

\bibitem[{{Komatsu} {et~al.}(2010){Komatsu}, {Smith}, {Dunkley}, {Bennett},
  {Gold}, {Hinshaw}, {Jarosik}, {Larson}, {Nolta}, {Page}, {Spergel},
  {Halpern}, {Hill}, {Kogut}, {Limon}, {Meyer}, {Odegard}, {Tucker}, {Weiland},
  {Wollack}, \& {Wright}}]{komatsu10}
{Komatsu}, E., {Smith}, K.~M., {Dunkley}, J., {Bennett}, C.~L., {Gold}, B.,
  {Hinshaw}, G., {Jarosik}, N., {Larson}, D., {Nolta}, M.~R., {Page}, L.,
  {Spergel}, D.~N., {Halpern}, M., {Hill}, R.~S., {Kogut}, A., {Limon}, M.,
  {Meyer}, S.~S., {Odegard}, N., {Tucker}, G.~S., {Weiland}, J.~L., {Wollack},
  E., \& {Wright}, E.~L. 2010, \apjs, submitted (arXiv:1001.4538)
  
\bibitem[Laing(1993)]{laing93} Laing, R.~A.\ 1993, Astrophysics 
and Space Science Library, 103, 95 
  
\bibitem[Laing et al.(1994)]{laing94} Laing, R.~A., Jenkins, 
C.~R., Wall, J.~V., \& Unger, S.~W.\ 1994, The Physics of Active Galaxies, 54, 201 


\bibitem[{{Leahy}(1993)}]{leahy93}
{Leahy}, J.~P. 1993, in Lecture Notes in Physics, Berlin Springer Verlag, Vol.
  421, Jets in Extragalactic Radio Sources, ed. {H.-J.~R{\"o}ser \&
  K.~Meisenheimer}, 1

\bibitem[Ledlow 
\& Owen(1995)]{ledlow95} Ledlow, M.~J. \& Owen, F.~N.\ 1995, \aj, 109, 853 

\bibitem[{{Ledlow} \& {Owen}(1996)}]{ledlow96}
{Ledlow}, M.~J. \& {Owen}, F.~N. 1996, \aj, 112, 9

\bibitem[{{Lin} \& {Mohr}(2007)}]{lin07}
{Lin}, Y.-T. \& {Mohr}, J.~J. 2007, \apjs, 170, 71

\bibitem[Lupton(1993)]{lupton93} Lupton, R.~H.\ 1993, Statistics in theory and practice (Princeton University Press)

\bibitem[{{Mackay}(1971)}]{mackay71}
{Mackay}, C.~D. 1971, \mnras, 154, 209

\bibitem[{{Mandelbaum} {et~al.}(2009){Mandelbaum}, {Li}, {Kauffmann}, \&
  {White}}]{mandelbaum09}
{Mandelbaum}, R., {Li}, C., {Kauffmann}, G., \& {White}, S.~D.~M. 2009, \mnras,
  393, 377

\bibitem[{{Narayan}(2005)}]{narayan05}
{Narayan}, R. 2005, \apss, 300, 177

\bibitem[{{Narayan} \& {Yi}(1995)}]{narayan95}
{Narayan}, R. \& {Yi}, I. 1995, \apj, 444, 231

\bibitem[{{O'Dea}(1998)}]{odea98}
{O'Dea}, C.~P. 1998, \pasp, 110, 493

\bibitem[{{Owen}(1993)}]{owen93}
{Owen}, F.~N. 1993, Lecture Notes in Physics, Berlin Springer Verlag, 421, 273

\bibitem[{{Owen} \& {Laing}(1989)}]{owen89}
{Owen}, F.~N. \& {Laing}, R.~A. 1989, \mnras, 238, 357

\bibitem[{{Owen} \& {White}(1991)}]{owen91}
{Owen}, F.~N. \& {White}, R.~A. 1991, \mnras, 249, 164

\bibitem[{{Petrosian}(1976)}]{petrosian76}
{Petrosian}, V. 1976, \apjl, 209, L1

\bibitem[{{Reyes} {et~al.}(2008){Reyes}, {Mandelbaum}, {Hirata}, {Bahcall}, \&
  {Seljak}}]{reyes08}
{Reyes}, R., {Mandelbaum}, R., {Hirata}, C., {Bahcall}, N., \& {Seljak}, U.
  2008, \mnras, 390, 1157

\bibitem[{{Salim} {et~al.}(2007){Salim}, {Rich}, {Charlot}, {Brinchmann},
  {Johnson}, {Schiminovich}, {Seibert}, {Mallery}, {Heckman}, {Forster},
  {Friedman}, {Martin}, {Morrissey}, {Neff}, {Small}, {Wyder}, {Bianchi},
  {Donas}, {Lee}, {Madore}, {Milliard}, {Szalay}, {Welsh}, \& {Yi}}]{salim07}
{Salim}, S., {Rich}, R.~M., {Charlot}, S., {Brinchmann}, J., {Johnson}, B.~D.,
  {Schiminovich}, D., {Seibert}, M., {Mallery}, R., {Heckman}, T.~M.,
  {Forster}, K., {Friedman}, P.~G., {Martin}, D.~C., {Morrissey}, P., {Neff},
  S.~G., {Small}, T., {Wyder}, T.~K., {Bianchi}, L., {Donas}, J., {Lee}, Y.,
  {Madore}, B.~F., {Milliard}, B., {Szalay}, A.~S., {Welsh}, B.~Y., \& {Yi},
  S.~K. 2007, \apjs, 173, 267

\bibitem[{{Shen} {et~al.}(2003){Shen}, {Mo}, {White}, {Blanton}, {Kauffmann},
  {Voges}, {Brinkmann}, \& {Csabai}}]{shen03}
{Shen}, S., {Mo}, H.~J., {White}, S.~D.~M., {Blanton}, M.~R., {Kauffmann}, G.,
  {Voges}, W., {Brinkmann}, J., \& {Csabai}, I. 2003, \mnras, 343, 978

\bibitem[{{Strauss} {et~al.}(2002){Strauss}, {Weinberg}, {Lupton}, {Narayanan},
  {Annis}, {Bernardi}, {Blanton}, {Burles}, {Connolly}, {Dalcanton}, {Doi},
  {Eisenstein}, {Frieman}, {Fukugita}, {Gunn}, {Ivezi{\'c}}, {Kent}, {Kim},
  {Knapp}, {Kron}, {Munn}, {Newberg}, {Nichol}, {Okamura}, {Quinn}, {Richmond},
  {Schlegel}, {Shimasaku}, {SubbaRao}, {Szalay}, {Vanden Berk}, {Vogeley},
  {Yanny}, {Yasuda}, {York}, \& {Zehavi}}]{strauss02}
{Strauss}, M.~A., et al.
  2002, \aj, 124, 1810

\bibitem[{{Tremaine} {et~al.}(2002){Tremaine}, {Gebhardt}, {Bender}, {Bower},
  {Dressler}, {Faber}, {Filippenko}, {Green}, {Grillmair}, {Ho}, {Kormendy},
  {Lauer}, {Magorrian}, {Pinkney}, \& {Richstone}}]{tremaine02}
{Tremaine}, S., {Gebhardt}, K., {Bender}, R., {Bower}, G., {Dressler}, A.,
  {Faber}, S.~M., {Filippenko}, A.~V., {Green}, R., {Grillmair}, C., {Ho},
  L.~C., {Kormendy}, J., {Lauer}, T.~R., {Magorrian}, J., {Pinkney}, J., \&
  {Richstone}, D. 2002, \apj, 574, 740

\bibitem[{{Urry} \& {Padovani}(1995)}]{urry95}
{Urry}, C.~M. \& {Padovani}, P. 1995, \pasp, 107, 803

\bibitem[{{Veilleux} \& {Osterbrock}(1987)}]{veilleux87}
{Veilleux}, S. \& {Osterbrock}, D.~E. 1987, \apjs, 63, 295

\bibitem[{{Willott} {et~al.}(2001){Willott}, {Rawlings}, {Blundell}, {Lacy}, \&
  {Eales}}]{willott01}
{Willott}, C.~J., {Rawlings}, S., {Blundell}, K.~M., {Lacy}, M., \& {Eales},
  S.~A. 2001, \mnras, 322, 536


\bibitem[Wing \& Blanton(2010)]{wing10} Wing, J.~D., \& Blanton, E.~L.\ 2010, \aj, submitted (arXiv:1008.1099) 




\bibitem[{{York} {et~al.}(2000){York}, {Adelman}, {Anderson}, {Anderson},
  {Annis}, {Bahcall}, {Bakken}, {Barkhouser}, {Bastian}, {Berman}, {Boroski},
  {Bracker}, {Briegel}, {Briggs}, {Brinkmann}, {Brunner}, {Burles}, {Carey},
  {Carr}, {Castander}, {Chen}, {Colestock}, {Connolly}, {Crocker}, {Csabai},
  {Czarapata}, {Davis}, {Doi}, {Dombeck}, {Eisenstein}, {Ellman}, {Elms},
  {Evans}, {Fan}, {Federwitz}, {Fiscelli}, {Friedman}, {Frieman}, {Fukugita},
  {Gillespie}, {Gunn}, {Gurbani}, {de Haas}, {Haldeman}, {Harris}, {Hayes},
  {Heckman}, {Hennessy}, {Hindsley}, {Holm}, {Holmgren}, {Huang}, {Hull},
  {Husby}, {Ichikawa}, {Ichikawa}, {Ivezi{\'c}}, {Kent}, {Kim}, {Kinney},
  {Klaene}, {Kleinman}, {Kleinman}, {Knapp}, {Korienek}, {Kron}, {Kunszt},
  {Lamb}, {Lee}, {Leger}, {Limmongkol}, {Lindenmeyer}, {Long}, {Loomis},
  {Loveday}, {Lucinio}, {Lupton}, {MacKinnon}, {Mannery}, {Mantsch}, {Margon},
  {McGehee}, {McKay}, {Meiksin}, {Merelli}, {Monet}, {Munn}, {Narayanan},
  {Nash}, {Neilsen}, {Neswold}, {Newberg}, {Nichol}, {Nicinski}, {Nonino},
  {Okada}, {Okamura}, {Ostriker}, {Owen}, {Pauls}, {Peoples}, {Peterson},
  {Petravick}, {Pier}, {Pope}, {Pordes}, {Prosapio}, {Rechenmacher}, {Quinn},
  {Richards}, {Richmond}, {Rivetta}, {Rockosi}, {Ruthmansdorfer}, {Sandford},
  {Schlegel}, {Schneider}, {Sekiguchi}, {Sergey}, {Shimasaku}, {Siegmund},
  {Smee}, {Smith}, {Snedden}, {Stone}, {Stoughton}, {Strauss}, {Stubbs},
  {SubbaRao}, {Szalay}, {Szapudi}, {Szokoly}, {Thakar}, {Tremonti}, {Tucker},
  {Uomoto}, {Vanden Berk}, {Vogeley}, {Waddell}, {Wang}, {Watanabe},
  {Weinberg}, {Yanny}, \& {Yasuda}}]{york00}
{York}, D.~G., et al.
  2000, \aj, 120, 1579

\bibitem[{{Zirbel}(1997)}]{zirbel97}
{Zirbel}, E.~L. 1997, \apj, 476, 489

\bibitem[{{Zirbel} \& {Baum}(1995)}]{zirbel95}
{Zirbel}, E.~L. \& {Baum}, S.~A. 1995, \apj, 448, 521

\end{thebibliography}

\end{document}